\documentclass{aa}  

\usepackage{graphicx}
\usepackage{txfonts}
\usepackage{mathtools,amssymb}
\usepackage{booktabs,amsmath,breqn}
\begin{document}

   \title{Characterizing the dust content of disk substructures in TW Hya}

   \author{E. Mac\'ias \inst{1,2}
   \and O. Guerra-Alvarado \inst{3}
   \and C. Carrasco-Gonz\'alez \inst{3}
   \and \'A. Ribas \inst{2}
   \and C. C. Espaillat \inst{4}
   \and J. Huang \inst{5}
   \and S. M. Andrews \inst{6}
          }

   \institute{Joint ALMA Observatory, Avenida Alonso de Córdova 3107, Vitacura, Santiago, Chile\\
              \email{enrique.macias@alma.cl}
         \and
        European Southern Observatory, Avenida Alonso de Córdova 3107, Vitacura, Santiago, Chile
        \and
        Instituto de Radioastronom\'ia y Astrof\'isica, UNAM, Apartado Postal 3-72, 58089 Morelia Michoac\'an, M\'exico
        \and 
        Institute for Astrophysical Research, Department of Astronomy, Boston University, 725 Commonwealth Avenue, Boston, MA 02215, USA
        \and
        NHFP Sagan Fellow, Department of Astronomy, University of Michigan, 323 West Hall, 1085 S. University Avenue, Ann Arbor, MI 48109, USA
        \and
        Harvard-Smithsonian Center for Astrophysics, 60 Garden Street, Cambridge, MA 02138, USA
              }

  \date{Received October 30, 2020; accepted February 8, 2021}

  \abstract
   {A key piece of information to understand the origin and role of protoplanetary disk substructures is their dust content. In particular, disk substructures associated with gas pressure bumps can work as dust traps, accumulating grains and reaching the necessary conditions to trigger the streaming instability.}
   {In order to shed some light on the origin and role that disk substructures play in planet formation, we aim to characterize the dust content of substructures in the disk of TW Hya.}
   {We present Atacama Large Millimeter Array (ALMA) observations of TW Hya at 3.1 mm with $\sim50$ milliarcsecond resolution. These new data were combined with archival high angular resolution ALMA observations at 0.87 mm, 1.3 mm, and 2.1 mm. We analyze these multiwavelength data to infer a disk radial profile of the dust surface density, maximum particle size, and slope of the particle size distribution.}
   {Most previously known annular substructures in the disk of TW Hya are resolved at the four wavelengths. Inside the inner 3 au cavity, the 2.1 mm and 3.1 mm images show a compact source of free-free emission, likely associated with an ionized jet. Our multiwavelength analysis of the dust emission shows that the maximum particle size in the disk of TW Hya is $>1$ mm. The inner 20 au are completely optically thick at all four bands, which results in the data tracing different disk heights at different wavelengths. Coupled with the effects of dust settling, this prevents the derivation of accurate density and grain size estimates in these regions. 
   At $r>20$ au, we find evidence of the accumulation of large dust particles at the position of the bright rings, indicating that these are working as dust traps. The total dust mass in the disk is between 250 and 330 $M_{\oplus}$, which represents a gas-to-dust mass ratio between 50 and 70. Our mass measurement is a factor of 4.5-5.9 higher than the mass that one would estimate using the typical assumptions of large demographic surveys.  }
   {Our results indicate that the ring substructures in TW Hya are ideal locations to trigger the streaming instability and form new generations of planetesimals.}


   \maketitle
%

\section{Introduction}

Dust grains are the building blocks of planetesimals. Through the sequential collision and aggregation of smaller particles, dust grains can grow from micron to millimeter sizes in a relatively efficient manner \citep{bir10}. However, theoretical models face the following two important bottlenecks to produce larger grains and planetesimals after this point in the dust growth process: the drift and fragmentation barriers. 

The drift barrier is the consequence of the decoupling of pebbles (millimeter- and centimeter-sized grains) from the gas, which results in the onset of a head wind onto the grains. This wind leads to a fast radial drift of pebbles toward the regions of high pressure at the center of the disk \citep{whi72}. The second obstacle, dubbed the fragmentation barrier, has its origin in the increasing fragmentation efficiency of pebbles as they grow, migrate inward, and experience higher relative velocities between them \citep{bir11}. As a consequence, 
models of grain growth cannot produce dust particles larger than $\sim10$ cm in the inner regions of the disk \citep{bir12}.
The combination of these two processes should produce a rapid decrease in the solid mass in disks within $<1$ Myr \citep{bra07,bir10}, fundamentally preventing the formation 
of planetesimal-sized objects, which is a necessary step in the planet formation process \citep{ray20}. Nevertheless, exoplanetary systems are found to be ubiquitous \citep{win15}.

The most likely solution to bypass these barriers and allow the formation of planetesimals can be summarized in two steps. The first one consists of the local concentration of pebbles in gas pressure maxima in the disk, which can halt or even stop their radial drift and reduce their relative velocities \citep{pin12}.
The local enhancement in the abundance of pebbles at these dust traps can then provide the ideal conditions to trigger the second step, the so-called streaming instability \citep{you05}. This instability arises due to the back-reaction force that drifting dust particles exercise on the gas, which amplifies the local pressure maxima, attracting more pebbles and creating a positive feedback loop. According to numerical simulations, the streaming instability can be triggered in protoplanetary disks as long as $\sim1-10$ mm particles are concentrated with local gas-to-dust mass ratios smaller than 25 \citep{car15,yan17}. This mechanism can result in the rapid formation of clumps of particles that can become self-gravitating, thus quickly forming kilometer-sized planetesimals from millimeter- and centimeter-sized particles \citep{sim16,abo19}. 

The streaming instability is now commonly accepted as a necessary step to form planetesimals in disks \citep[][and references therein]{ray20}. A key to understand planet formation is thus to find in which environments, and at which locations in the disk, the physical conditions suitable to trigger the streaming instability can be met; that is to say, where, when, and how millimeter- and centimeter-sized dust particles can be accumulated in protoplanetary disks.

In recent years, observations taken with the Atacama Large Millimeter/submillimeter Array (ALMA) have revealed significant levels of substructure in disks. Rings, gaps, and azimuthal asymmetries are detected in protoplanetary disks across different evolutionary stages \citep[e.g.,][]{alm15,and16,per16,don18}. Even though all high resolution ALMA surveys up to date present an inherent bias toward large and bright disks, the results so far indicate that disk substructures are likely ubiquitous \citep[e.g.,][]{and18a,lon18}. This discovery has been transformational, 
partly because these substructures could represent the preferential locations to trap pebbles and form new generations of planetesimals. 
A number of physical mechanisms have been proposed to explain the observed substructures \citep[e.g.,][]{joh09,oku16,bai17,bae17}. While some of these scenarios do not require the onset of a dust trap at a pressure bump, many observational studies have found evidence of dust trapping in substructures \citep[e.g.,][]{van16,dul18,mac18,pin18,car19,mac19}.
In order to understand if these substructures can harbor the necessary conditions to trigger the streaming instability, it is key to study the particle size distribution at these positions.

Multiwavelength observations at (sub)millimeter wavelengths are one of the most efficient tools to study the particle size distribution in disks. This observational diagnosis relies on the relationship between the frequency dependency of the dust opacity and the dust particle size distribution. If the dust emission is optically thin, this relationship becomes relatively simple in the Rayleigh-Jeans regime: the spectral index of the (sub)millimeter dust emission is equal to $2+\beta$, where $\beta$ is the spectral index of the dust opacity \citep{bec90} and lower values of $\beta$ indicate larger dust particles \citep{miy93,dal01}. 
However, high optical depths also result in flatter spectral indices, which can make it challenging to distinguish between optically thick emission or the presence of large dust grains \citep{ric12}.
This limitation has been faced by a number of studies analyzing dual-band observations of disks, which tend to display lower values of the spectral index at the rings compared to those at the gaps or cavities \citep[e.g.,][]{pin15,pin17a,caz18,hua18a,hua20,lon20}.
In order to resolve this dichotomy and better characterize the dust content in disk substructures, a detailed analysis of observations at multiple wavelengths is required. 

So far, few sources have been studied with high spatial resolution at more than two wavelengths \citep{liu17,car19,mac19}. Although these studies found evidence of the presence of local accumulations of millimeter- and centimeter-sized particles, conflicting results have arisen from an alternative observational diagnosis: ALMA polarization observations. The polarization pattern and high polarization fraction detected in a number of disks has been explained as being produced by dust self-scattering with a maximum grain size of $\sim100~\mu$m, which is in dire contrast with the millimeter and centimeter sizes found by analyzing the (sub)millimeter spectrum of the dust emission \citep{kat16a,kat16b,kat17,hul18,den19}. More recently, it has been proposed that these polarization results could be explained by larger yet nonspherical particles \citep{kir20}. While more work is required to fully understand the polarization of dust continuum emission, more detailed multiwavelength studies are also needed to place more robust constraints on the particle size distribution in protoplanetary disks and within their substructures.

Here we focus on TW Hya (SpT=M0.5, $M_{\star}\sim0.6~M_{\odot}$, age$\sim8$ Myr; \citealp{sok18}), a young stellar object surrounded by one of the most studied circumstellar disks \citep[e.g.,][]{cal02,wil05,hug07,and16}. Despite its relatively advanced age, the disk still harbors a significant amount of gas \citep{ber13}. It is among the closest  protoplanetary disks to Earth ($d=59.5$; \citealp{bai18,gai18}) which, together with its almost face-on inclination ($i\sim5^{\circ}$; \citealp{and12}), makes TW Hya one of the best laboratories to study disk evolution and planetary formation.

High angular resolution ALMA observations of TW Hya at 0.87 mm, 1.3 mm, and 2 mm have revealed multiple gaps in the disk from 1 to 47 au \citep{nom16,and16,tsu16}. At least one of the gaps, at $20-30$ au, is consistent with a planetary origin \citep{nom16}, while the inner cavity at $\sim1$ au could be the result of the formation of a young super-Earth \citep{and16}, or the onset of disk photoevaporation \citep{erc17}. A small clump of 1.3 mm emission was recently detected at $\sim52$ au from the star \citep{tsu19}. This emission was interpreted to potentially trace a small vortex or circumplanetary material around a Neptune-mass planet. More recently, \citet{nay20} suggested that a planet located within this clump could be releasing significant amounts of gas and dust to the disk,
which could explain why so much material still surrounds such an old system. On the other hand, \citet{van21} suggested that a vortex is the most likely explanation by comparing its location with the asymmetries detected in other protoplanetary disks.

The spectral index radial profile of TW Hya at 0.87-1.3 mm and 1.3-2 mm shows higher values at the position of the gaps, as well as values close to 2 at the rings \citep{tsu16,hua18a}. Individually, these results were interpreted to be consistent with either optically thick emission or optically thin emission from millimeter- and centimeter-sized dust grains. More recently, \citet{ued20} analyzed multiwavelength observations of the inner 10 au of TW Hya, finding extremely high dust surface densities as well as relatively low maximum grain sizes of $\sim300~\mu$m.

In this study we present new $\sim50$ milliarcsecond observations at 3.1 mm of the disk around TW Hya. These data represent the longest wavelength at which high sensitivity and high spatial resolution observations have been obtained for TW Hya, providing a detailed view of the disk substructures at a lower optical depth than previous ALMA observations. We analyze these new data together with archival high resolution observations at 0.87 mm, 1.3 mm, and 2.1 mm. In order to shed some light on the origin of the disk substructures in TW Hya, as well as on the role that they play in the planet formation process, we use these multiwavelength data to robustly characterize the dust content of the disk and its substructures. The article is structured as follows. In section \ref{sec:obs} we describe the new 3.1 mm observations, together with the data reduction process of these and the archival data. Section \ref{sec:results} explains the main observational results obtained from these data. In section \ref{sec:mod} we describe our modeling of the multiwavelength observations, aimed at characterizing the dust content of the disk. Section \ref{sec:discussion} discusses the main results and implications of our analysis. Finally, section \ref{sec:conclusion} summarizes our findings and conclusions.

\section{Observations}\label{sec:obs}

The new ALMA long baseline observations at 3 mm (Band 3) were obtained during Cycle 6 in five different executions carried out from 2019 June 24 to 2019 July 08 (project code: 2018.1.01218.S). Additionally, we retrieved lower resolution archival data at 3 mm to cover the shorter baselines, as well as archival data at Bands 7, 6, and 4. Details about the dates, on-source time, antenna configuration, correlator setup, and flux calibrator of all the used datasets can be found in Table \ref{tab:obs}. When retrieving archival data, we used only spectral windows with bandwidth $\sim2$ GHz (i.e., we avoided using narrow spectral windows dedicated to line observations), and with central frequencies within $\sim0.5$ GHz of the central frequency of the other datasets in the band to avoid issues when combining them. The only exception to this was the combination of the long baseline data at Band 7 (project code: 2015.1.00686.S) with the other Band 7 datasets.

\begin{table*}
\caption{\label{tab:obs}Summary of ALMA Observations.}
\centering
\resizebox{\textwidth}{!}{
\begin{tabular}{cccccccc}
\hline\hline
Project & P.I. & Date &  On-source & Baselines & Frequencies & Flux & CASA    \\
Code    &      &      & time (min) &    (m)    &    (GHz)\tablefootmark{a}    & cal. & version\tablefootmark{b} \\
\hline
\multicolumn{8}{c}{Band 7} \\
\hline
2015.1.00686.S & S. Andrews & 2015 Nov 23 & 42.5 & 23 -- 14321 & 344.5, 345.8, 355.1, 357.1 & J1307-2934 & 4.5.0 \\
 &  & 2015 Nov 30 & 43.9 & 27 -- 10804 & 344.5, 345.8, 355.1, 357.1 & J1107-4449 & 4.5.0 \\
 &  & 2015 Dec 01 & 43.7 & 17 -- 10804 & 344.5, 345.8, 355.1, 357.1 & J1107-4449 & 4.5.0 \\
2016.1.00629.S  & I. Cleeves & 2016 Dec 30 & 33.4 & 15 -- 460 & 343.0 & J1037-2934 & 4.7.2 \\
  &  & 2016 Dec 30 & 21.7 & 15 -- 460 & 343.0 & J1058+0133 & 4.7.2 \\
  &  & 2017 Jul 04 & 44.9 & 21 -- 2647 & 343.0 & J1037-2934 & 4.7.2 \\
  &  & 2017 Jul 09 & 45.2 & 17 -- 2647 & 343.0 & J1037-2934 & 4.7.2 \\
  &  & 2017 Jul 14 & 45.7 & 19 -- 1458 & 343.0 & J1037-2934 & 4.7.2 \\
  &  & 2017 Jul 20 & 44.8 & 17 -- 3697 & 343.0 & J1037-2934 & 4.7.2 \\
  &  & 2017 Jul 21 & 44.8 & 17 -- 3697 & 343.0 & J1256-0547 & 4.7.2 \\
2016.1.00173.S  & T. Muto & 2018 Jan 23 & 34.4 & 15 -- 1398 & 336.5, 338.4, 348.5, 350.5 & J1107-4449 & 5.4.0 \\
  &  & 2018 Jan 23 & 34.4 & 15 -- 1398 & 336.5, 338.4, 348.5, 350.5 & J1107-4449 & 5.4.0 \\
\hline
\multicolumn{8}{c}{Band 6} \\
\hline
2015.A.00005.S & T. Tsukagoshi & 2015 Dec 01 & 40.2 & 17 -- 10804 & 224.0, 226.0, 240.0, 242.0 & J1037-2934 & 4.5.0 \\
2016.1.00842.S & T. Tsukagoshi & 2017 May 15 & 10.8 & 15 -- 1121 & 224.0, 226.0, 240.0, 242.0 & J1107-4449 & 4.7.2 \\
2017.1.00520.S & T. Tsukagoshi & 2017 Nov 20 & 42.5 & 92 -- 8548 & 224.0, 226.0, 240.0, 242.0 & J1107-4449 & 5.1.1 \\
 &  & 2017 Nov 23 & 43.1 & 92 -- 8548 & 224.0, 226.0, 240.0, 242.0 & J1058+0133 & 5.1.1 \\
 &  & 2017 Nov 25 & 42.1 & 92 -- 8548 & 224.0, 226.0, 240.0, 242.0 & J1037-2934 & 5.1.1 \\
\hline
\multicolumn{8}{c}{Band 4} \\
\hline
2015.A.00005.S & T. Tsukagoshi & 2015 Dec 02 & 43.3 & 17 -- 10804 & 138.0, 140.0, 150.0, 152.0 & J1307-2934 & 4.5.0 \\
2015.1.00845.S & C. Favre & 2016 Apr 29 & 41.2 & 15 -- 639 & 145.0 & J1107-4449 & 4.5.3 \\
 &  & 2016 Apr 29 & 41.2 & 15 -- 639 & 145.0 & J1037-2934 & 4.5.3 \\
2016.1.00842.S & T. Tsukagoshi & 2016 Oct 21 & 11.8 & 18 -- 1808 & 138.0, 140.0, 150.0, 152.0 & J1307-2934 & 4.7.2 \\
 &  & 2017 Sep 28 & 39.1 & 41 -- 14851 & 138.0, 140.0, 150.0, 152.0 & J1107-4449 & 4.7.2 \\
 &  & 2017 Dec 05 & 11.8 & 41 -- 3637 & 138.0, 140.0, 150.0, 152.0 & J1037-2934 & 4.7.2 \\
\hline
\multicolumn{8}{c}{Band 3} \\
\hline
2016.1.00229.S & E. Bergin & 2017 Aug 01 & 43.0 & 17 -- 3108 & 97.8 & J1037-2934 & 4.7.2 \\
2018.1.01218.S & E. Mac\'ias & 2019 Jun 24 & 43.5 & 83 -- 16196 & 90.5, 92.5, 102.5, 104.5 & J1037-2934 & 5.4.0 \\
 &  & 2019 Jun 27 & 43.5 & 83 -- 14969 & 90.5, 92.5, 102.5, 104.5 & J1307-2934 & 5.4.0 \\
 &  & 2019 Jul 04 & 43.6 & 83 -- 16196 & 90.5, 92.5, 102.5, 104.5 & J1307-2934 & 5.4.0 \\
 &  & 2019 Jul 07 & 43.7 & 149 -- 13894 & 90.5, 92.5, 102.5, 104.5 & J1307-2934 & 5.4.0 \\
 &  & 2019 Jul 08 & 43.5 & 149 -- 13894 & 90.5, 92.5, 102.5, 104.5 & J1307-2934 & 5.4.0 \\
\hline
\end{tabular}}
\tablefoot{
\tablefoottext{a}{Central frequency of spectral windows used for continuum imaging.}
\tablefoottext{b}{CASA version used for data calibration.}
}
\end{table*}

Data calibration was performed using the calibration pipeline and scripts developed by ALMA staff. The last column in table \ref{tab:obs} shows the version of \texttt{CASA} (Common Astronomy Software Applications; \citealp{mcm07}) used for the calibration of each dataset. For further data processing and cleaning, \texttt{CASA} version 5.6.1 was used. The data for each execution were revised, and additional flagging was applied when necessary. Since we are only interested in the continuum emission, we performed channel averaging to each dataset up to a width $\sim125$ MHz. 

Before combining data from different executions (within a project or from different projects), a few additional steps were performed to ensure that the data were accurately centered in position (i.e., correcting from proper motions and astrometric errors), and to minimize the absolute flux calibration systematic errors. The first step was to shift the phase center of the visibilities of each execution to the position of TW Hya at the date of the observations, as determined from the position and proper motions reported by Gaia DR2 \citep{gai18}, using the task \texttt{fixvis} in \texttt{CASA}. Then, the \texttt{table} tool in \texttt{CASA} was used to modify the coordinates of the field to align the different executions\footnote{This step is sometimes performed using the task \texttt{fixplanets} in \texttt{CASA}. However, this task only works if the data are in the FK5 celestial reference system. In order to keep our data in the ICRS system we used the \texttt{table} tool to manually change the coordinates of the field in the measurement sets.}. Once the coordinates of each dataset were centered and aligned, we rescaled the fluxes following a similar procedure to \citet{and18a}: we deprojected the real part of the visibilities, selected a range of uv-distances where the uv coverage of two datasets overlaps (with an upper limit of 800 $k\lambda$ to avoid decorrelation issues), binned the visibilities in that range, and compared them to obtain a rescaling factor. One of the two datasets was then rescaled using the task \texttt{gencal}. The reference dataset at each band was selected based on the flux calibrator used, the time between the observation and the last flux monitoring of that calibrator, and the behavior of the calibrator during that time. After rescaling the flux, the different executions were combined into a single dataset. 

In order to improve the quality of the data and further correct for small phase shifts produced by astrometric errors, the data were then self-calibrated. Imaging during self-calibration was performed using the task \texttt{tclean} with the \texttt{mtmfs} deconvolver \citep{rau11}, assuming that the frequency dependency of the emission follows a Taylor expansion to quadratic terms (i.e., $nterms=2$),  and \texttt{briggs} weighting with $robust=0.5$. For datasets with angular resolution poorer than $\sim1''$, a single point-source scale was used ($scales=0$). Datasets with higher resolution resolve the disk emission, so various scales at 0 (point-source), 1, 3, and 5 times the beam size were used. Phase self-calibration was performed first, in an iterative process starting with longer time solution intervals and decreasing them until no improvement was seen in the peak signal-to-noise ratio (S/N) or until reaching the integration time of the data. Then, amplitude self-calibration was attempted, although it only provided improvements in the S/N in a few cases with low resolution and very high S/N. In general, self-calibration on individual datasets yielded improvements in S/N ranging from $50\%$ to $300\%$.

Some projects had two different scheduling blocks using different antenna configurations to cover short and long baselines. For these projects each scheduling block was first self-calibrated separately after combining the multiple executions of each scheduling block. The data were then checked for any possible large position offset between the different scheduling blocks. The nominal absolute astrometric accuracy of ALMA is $\sim10\%$ the beam size, so datasets with different angular resolutions could have relatively large position offsets. These position mismatches need to be fixed before self-calibrating the combined datasets. If a significant offset was found (offsets larger than $1/3$ the extended configuration beam size), it was corrected by measuring the position of the emission with Gaussian fits to the images and then aligning the visibilities. Afterward, the short and long baseline data were combined and self-calibrated. 

A similar approach was followed when combining data from different projects: the coordinates were shifted and aligned to correct for proper motions, the flux was rescaled, position offsets were corrected if present, and finally the data were combined and self-calibrated together. The data and self-calibration solutions were checked after each step to verify that self-calibration did correct for small shifts in position. Overall, we found that self-calibration yielded larger improvements in the signal-to-noise ratio when performed on the combined datasets than on the high resolution datasets alone, thanks to the better quality model obtained during the cleaning process. A final shift was applied to the self-calibrated combined data to ensure that the data at the different bands were centered at the same position.

Once the final self-calibrated combined measurement sets were obtained, the \texttt{CASA} task \texttt{statwt} was used to recompute the weights. Finally, the data were imaged using the task \texttt{tclean}. The \texttt{mtmfs} deconvolver was employed, setting $nterms=2$, and using multiple scales at 0 (point-source), 1, 3, and 5 times the beam size. The pixel size was set to $\sim0.1$ times the beam size. Different values of the $robust$ parameter were explored to find a balance between angular resolution and sensitivity. The original synthesized beam shape of the Band 4 data was quite elongated, so we also explored different uv-taper values to get a more circular beam shape. Additionally, higher resolution maps at each band were produced to showcase the inner regions of the disk in more detail.
The final values of $robust$ used, together with the resulting synthesized beam sizes, rms values, peak S/N, and integrated flux densities can be found in Table \ref{tab:obsres}. The flux densities were obtained by integrating the emission within a $2''$ radius circle. The rms was measured within an annulus with $2''$ of inner radius and $4''$ of outer radius. 

\begin{table*}
\caption{\label{tab:obsres}Image processing and results.}
\centering
\resizebox{\textwidth}{!}{
\begin{tabular}{cccccccc}
\hline\hline
Band & Central  & Wavelength & robust & Beam shape & rms & Peak & Flux density \\
 & freq. (GHz) & (mm) &  &  & ($\mu$Jy beam$^{-1}$) & S/N & (mJy)\tablefootmark{a} \\
\hline
7 & 346.8 & 0.865 & -0.2 & $0\rlap.''036\times0\rlap.''028$, PA$=73^{\circ}$ & 26 & 90 & $1542\pm3$ ($\pm150$) \\
- & 346.8 & 0.865 & -1.0 & $0\rlap.''023\times0\rlap.''018$, PA$=86^{\circ}$ & 48 & 24 & $1517\pm8$ ($\pm150$) \\
6 & 233.0 & 1.29 & 0.0 & $0\rlap.''036\times0\rlap.''031$, PA$=-85^{\circ}$ & 9.5 & 135 & $575.4\pm0.9$ ($\pm30$) \\
- & 233.0 & 1.29 & -1.0 & $0\rlap.''029\times0\rlap.''025$, PA$=-84^{\circ}$ & 20 & 45 & $511\pm3$ ($\pm30$) \\
4 & 145.0 & 2.07 & -0.6\tablefootmark{b} & $0\rlap.''046\times0\rlap.''039$, PA$=46^{\circ}$ & 12 & 62 & $151.1\pm0.9$ ($\pm8$)\\
- & 145.0 & 2.07 & -1.0 & $0\rlap.''045\times0\rlap.''026$, PA$=46^{\circ}$ & 17 & 34 & $143.8\pm1.7$ ($\pm8$) \\
3 & 97.5 & 3.08 & 0.5 & $0\rlap.''052\times0\rlap.''050$, PA$=-10^{\circ}$ & 4.5 & 164 & $59.8\pm0.3$ ($\pm3$)\\
- & 97.5 & 3.08 & -1.0 & $0\rlap.''034\times0\rlap.''033$, PA$=74^{\circ}$ & 13 & 43 & $42.3\pm1.3$ ($\pm3$) \\
\hline
\end{tabular}}
\tablefoot{
\tablefoottext{a}{Uncertainties in parenthesis include the nominal $10\%$ (for Band 7) and $5\%$ (for Bands 6, 4, and 3) absolute calibration uncertainties for ALMA.}
\tablefoottext{b}{A robust value of $-0.6$ together with uv-tapering of $0\rlap.''03\times0\rlap.''001$ (PA$=136^{\circ}$) was used to produce a more circular beam.}
}
\end{table*}

Furthermore, spectral index ($\alpha$) maps were obtained combining data at multiple bands in the task \texttt{tclean}. This allowed us to obtain spectral index maps between the four bands. The deconvolver algorithm \texttt{mtmfs} was employed for this purpose, using \texttt{nterms=2}. Table \ref{tab:obsresspind} shows more details about the cleaning process as well as the resulting beam shapes
of the resulting images.

\begin{table*}
\caption{\label{tab:obsresspind} Image processing and results for spectral index maps.}
\centering
\begin{tabular}{ccccc}
\hline\hline
Band & Central  & Wavelength & robust & Beam shape \\
 & freq. (GHz) & (mm) &  &  \\
\hline
7+6 & 290.5 & 1.03 & 0.0 & $0\rlap.''037\times0\rlap.''031$, PA$=-89^{\circ}$ \\
6+4 & 190.0 & 1.58 & 0.2 & $0\rlap.''043\times0\rlap.''036$, PA$=64^{\circ}$ \\
4+3 & 121.3 & 2.47 & 0.3 & $0\rlap.''048\times0\rlap.''044$, PA$=39^{\circ}$ \\
\hline
\end{tabular}
\end{table*}

\section{Observational results}\label{sec:results}

\begin{figure*}
\centering
\includegraphics[width=\textwidth]{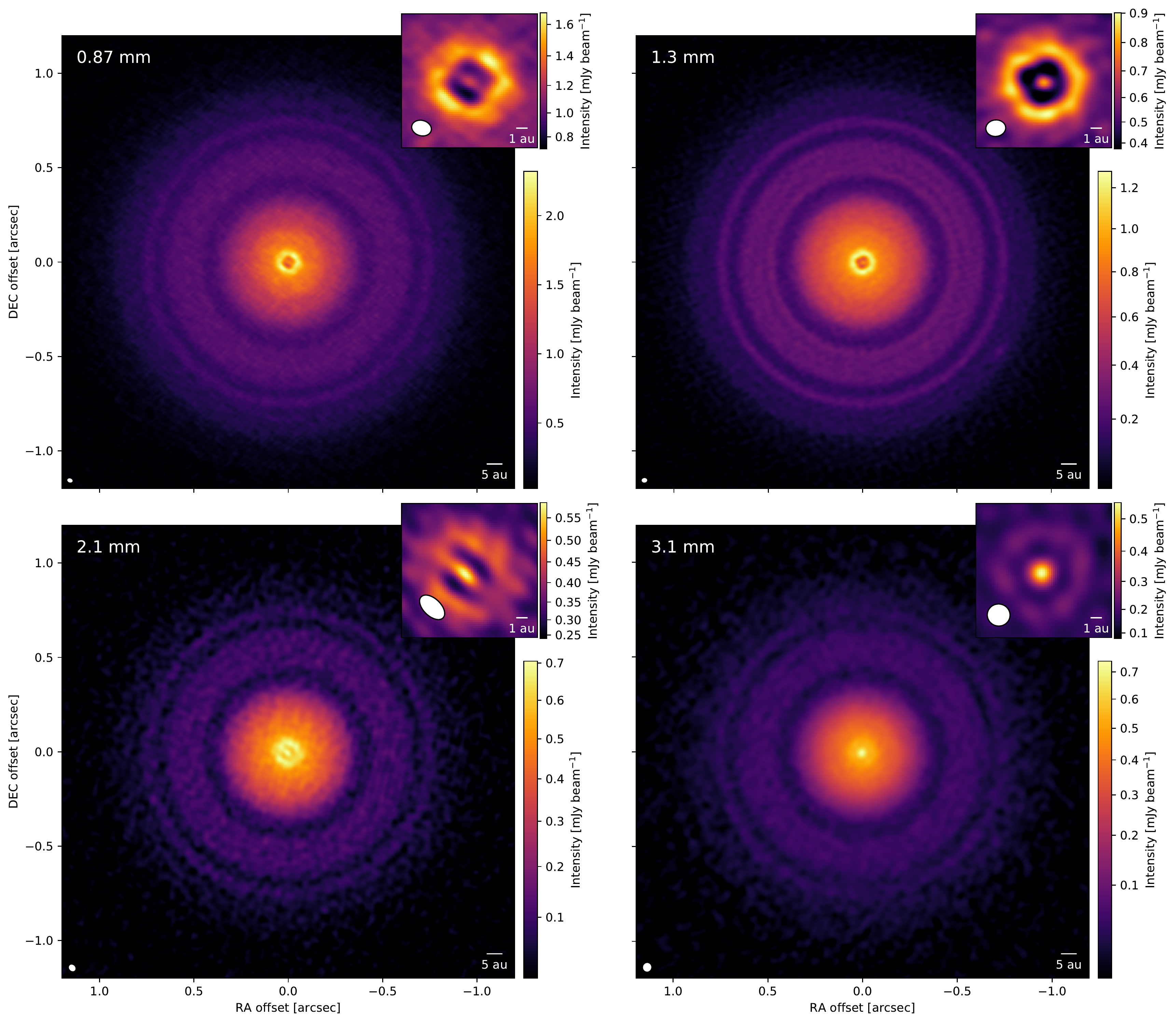}
\caption{Images of the dust thermal emission of the protoplanetary disk around TW Hya at 0.87 mm (top-left), 1.3 mm (top-right), 2.1 mm (bottom-left), and 3.1 mm (bottom-right). The inset in each panel shows a zoom of the inner $0\rlap.''2$ region of the disk (i.e., 12 au), using an image with higher angular resolution. The rms, beam sizes, and flux densities of all images are listed in Table \ref{tab:obsres}.}
\label{fig:maps}%
\end{figure*}

The obtained images at 0.87 mm (Band 7), 1.3 mm (Band 6), 2.1 mm (Band 4), and 3.1 mm (Band 3) are shown in Figure \ref{fig:maps}, together with a higher resolution inset of the inner regions. A summary of the properties of these two images is listed in Table \ref{tab:obsres}. Our images clearly detect and resolve at the different wavelengths most substructures previously detected in TW Hya. At Bands 7 and 6, our images are similar to the ones presented by \citet{hua18a} and \citet{tsu19}, respectively, although with higher resolution at Band 6, and higher sensitivity at Band 7. Some of the Band 4 data used in this paper were also first presented in \citet{tsu16}, but here we combined those observations with other archival data to obtain higher resolution and higher sensitivity images.

The 1.3 mm image, in particular, shows an extreme degree of detail, clearly resolving several concentric rings with varying contrasts. At Bands 4 and 3 we resolve most of the substructures at $r<44$ au, albeit the lower brightness of the dust thermal emission at these wavelengths results in lower S/N. Nevertheless, the high resolution of our images allows us to resolve the substructures including the inner cavity at $\sim3$ au first reported by \citet{and16}. The clump of emission previously found by \citet{tsu19} at $\sim52$ au is clearly detected to the southwest of the disk in our 0.87 mm and 1.3 mm images. Its presence is also evident at 3.1 mm. In this study we focus mainly on the radial intensity profile of the disk, and a more in depth analysis of the feature at 52 au is deferred to a future paper.

\begin{figure*}
\centering
\includegraphics[width=0.95\textwidth]{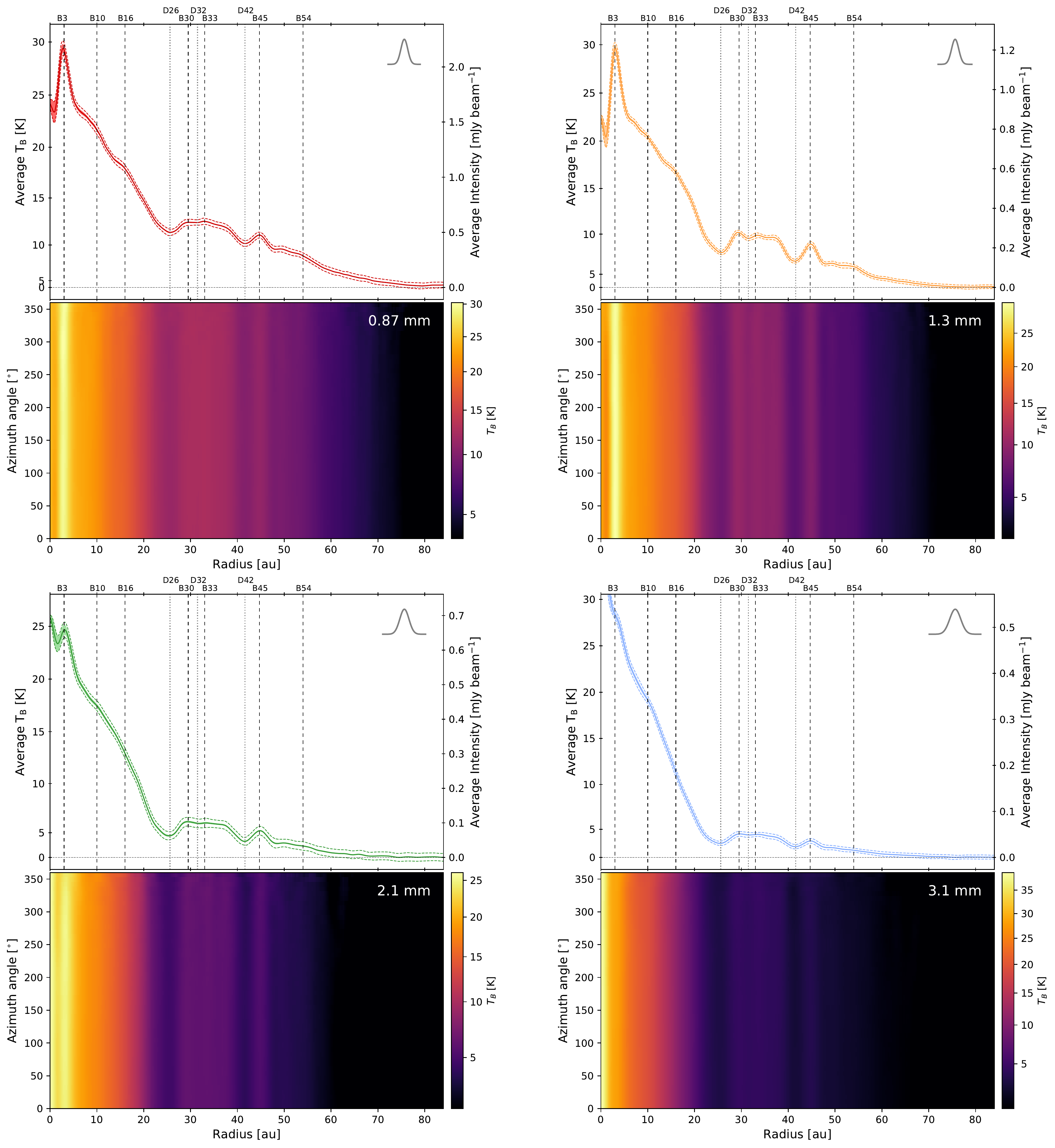}
\caption{Deprojected images (bottom half of each panel) and azimuthally averaged radial intensity profile (top half of each panel) of the disk of TW Hya at 0.87 mm (top-left), 1.3 mm (top-right), 2.1 mm (bottom-left), and 3.1 mm (bottom-right). The shaded color regions in each profile indicate the error of the mean at each radius, while the colored dashed lines show the standard deviation for comparison. The vertical lines mark the positions of the bright (dashed) and dark (dotted) rings in the disk. The top right insets in the radial profile panels show the geometric mean of the beam size of each image. We note that the vertical axis scale in the radial profile at 3.1 mm is intentionally cut to better display the disk emission. A better view of the disk inner region at this wavelength is shown in Fig. \ref{fig:highresprofiles}.}
\label{fig:profiles}%
\end{figure*}

\begin{figure*}
\centering
\includegraphics[width=\textwidth]{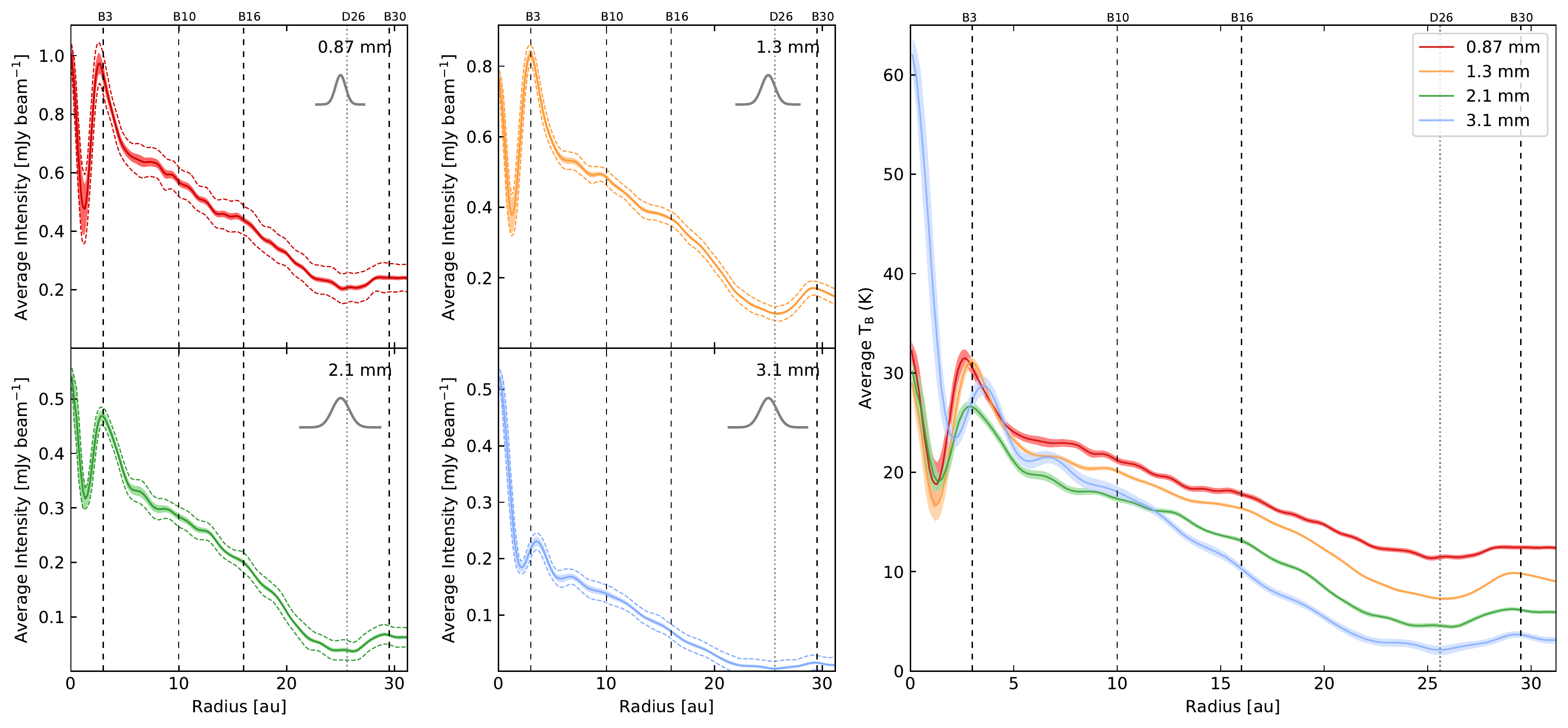}
\caption{Azimuthally averaged radial intensity profile of the inner $30$ au of the disk of TW Hya at 0.87 mm (top-left), 1.3 mm (top-middle), 2.1 mm (bottom-left), and 3.1 mm (bottom-middle), obtained from the higher resolution images shown in the insets of Fig. \ref{fig:maps}. In order to minimize the effects of the elongated beam at 2.1 mm, the radial profile at this wavelength was obtained by averaging the emission only within two $90^{\circ}$ slices in PA across the minor axis of the beam. The inset in this panel shows the geometric mean of the full beam size, so the resolution of the profile will be slightly better than the one displayed by the inset. The lines and colored regions indicate the same as in Fig.\ref{fig:profiles}. The right panel displays the same four profiles in units of brightness temperature.}
\label{fig:highresprofiles}%
\end{figure*}

From these images we obtained the deprojected images and azimuthally averaged radial intensity profiles shown in Figure \ref{fig:profiles}. The deprojection was performed using an inclination of $5^{\circ}$ and a position angle of $152^{\circ}$ \citep{hua18a}. The radial intensity profiles were obtained by averaging the emission in concentric ellipses with this same inclination and position angle. The uncertainty of the radial profiles is computed as the error of the mean at each radius, considering the beam size as the smallest independent unit of area:
\begin{equation}
    \sigma_{\bar{I}} = \frac{\sigma_i}{\sqrt{N_B}} = \frac{\sigma_i}{\sqrt{A_i/A_{beam}}},
\end{equation}
where $\sigma_i$ is the standard deviation within the concentric ellipse, $N_B$ is the number of beams within the ellipse, $A_i$ is the area of the ellipse, and $A_{beam}$ is the area of the beam. For comparison, the standard deviation at each radius is also potted in the radial profiles in Figure \ref{fig:profiles}. This figure also shows the position of some of the bright  (i.e., "rings") and dark (i.e., "gaps") annular features in the disk listed by \citet{hua18b}. We follow the naming convention from DSHARP \citep{hua18b} and label the rings as "B" or "D" (depending on whether they show a bright or dark annulus) followed by their position in au. Despite not being completely resolved, we indicate the position of the edge of two "plateau"-like features at $\sim7$ au and $\sim16$ au. Furthermore, we also indicate the position of the steep drop-off of emission at $\sim54$ au. Following the naming criterion for "plateau"-like features by \citet{hua18b}, we label these three features as B7, B16, and B54. Similar radial intensity profiles with higher resolution are plotted in Figure \ref{fig:highresprofiles}. These profiles are obtained from the higher resolution images shown in the insets of Fig. \ref{fig:maps}, and they focus only on the inner $30$ au of the disk.

\subsection{Spectral index profiles}

\begin{figure*}
\centering
\includegraphics[width=\textwidth]{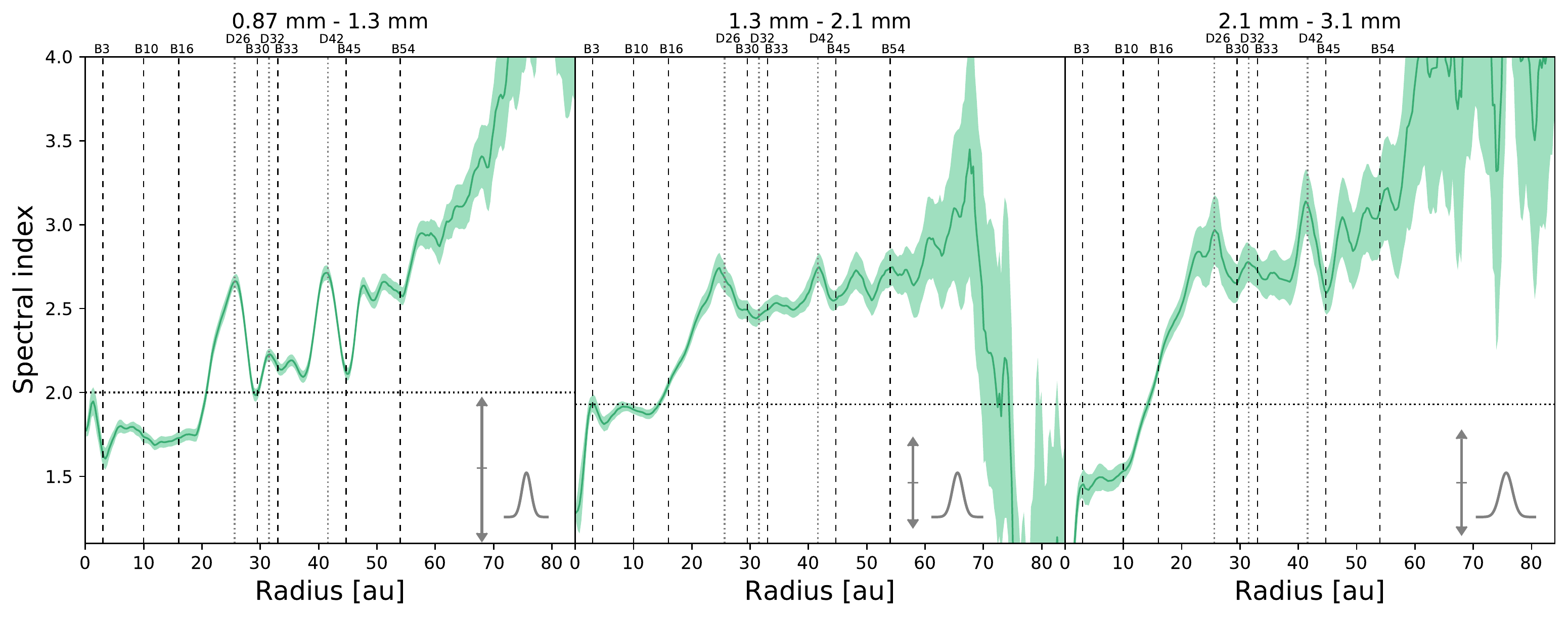}
\caption{Spectral index radial profiles of the dust emission of the disk around TW Hya between 0.87 mm and 1.3 mm (left), 1.3 mm and 2.1 mm (middle), and 2.1 mm and 3.1 mm (right). These radial profiles were obtained by azimuthally averaging the spectral index maps obtained with the multiband imaging in \texttt{tclean} (see Sect. \ref{sec:obs}). The shaded colored regions indicate the error of the mean at each radii, computed as in the averaged radial intensity profiles. The Gaussian insets in the lower-right corner of each plot show the geometric mean of the beam size. The gray error bars in the lower-right corner of each plot indicate the maximum possible vertical shift of the spectral index profiles introduced by flux calibration systematic errors. We note that these shifts would affect all radii simultaneously, so they do not affect the relative shape of the spectral index profiles. The vertical dashed and dotted lines are the same as in Fig.\ref{fig:profiles}.}
\label{fig:spindex}%
\end{figure*}

We obtained spectral index maps combining data at multiple bands in \texttt{tclean}. Figure \ref{fig:spindex} shows the spectral index profiles obtained from these maps. The solid line in each panel shows the azimuthally averaged value of the spectral index, with the shaded colored region indicating the error of the mean. We note that, due to the absolute flux calibration uncertainty, a systematic error could be present in these maps, increasing or decreasing the values at all radii. The gray error bars in each panel of this Figure indicate the maximum potential shift in the spectral index profile introduced by these systematic errors (assuming the nominal errors of 10\% at Band 7, and 5\% at Bands 6, 4, and 3). This shift, however, will not affect the relative changes with radius seen in the individual profiles.

Our spectral index profiles at 0.87-1.3 mm and 1.3-2.1 mm are similar to those obtained by \citet{hua18a} and \citet{tsu16}, respectively. The three spectral index profiles show some common features: a flat region with $\alpha<2$ in the inner disk ($r\lesssim20$ au), larger values at the position of the two main gaps (D25 and D41), lower values at the position of the brighter rings (e.g., B29 and B44), and an overall increasing trend with radius (Figure \ref{fig:spindex}). 
We note that our 1.3-2.1 mm and 2.1-3.1 mm spectral index profiles also show a sharp decrease at $r<3$ au, but these low values are due to the free-free emission contributing significantly at 2.1 and 3.1 mm (see Sect. \ref{subsec:freefree}). 

The $\alpha<2$ region at $r\lesssim20$ au has been attributed to optically thick emission with high albedo (i.e., dominated by self-scattering; \citealp{ued20}). The fact that the 2.1-3.1 mm spectral index in this region is still below 2 shows that even in this range of wavelengths the innermost disk is optically thick. However, the extent of the region with $\alpha<2$ does change with wavelength. At 0.87-1.3 mm the flat region extends up to $r\sim20$ au (Figure \ref{fig:spindex}, left), while it only extends up to $\sim15$ au at 1.3-2.1 mm (Figure \ref{fig:spindex}, middle), and up to $\sim10$ au at 2.1-3.1 mm (Figure \ref{fig:spindex}, right). Another notable difference in the spectral index profile between the different wavelength ranges is the contrast between the spectral index in the inner regions and the rest of the disk. This can be easily seen by comparing the spectral index at $r\sim10$ au with the value at $r\sim35$ au (at the group of rings between D26 and D42). The difference between the spectral index at these two positions is $\sim0.4$ at 0.87-1.3 mm, $\sim0.6$ at 1.3-2.1 mm, and $\sim1.3$ at 2.1-3.1 mm. \footnote{We note that the difference between the spectral index at two different radii is not affected by the flux calibration error.}.

These changes in the spectral index with wavelength likely indicate that, despite the innermost region of the disk remaining optically thick at all bands, there is an overall decrease in optical depth at longer wavelengths. These frequency variations in dust optical depth can be used to estimate the properties of the dust emission, namely, the dust density, dust temperature, and dust opacity. We analyze the frequency dependency of the dust emission in more detail in section Sect. \ref{sec:mod}.

\subsection{Central compact emission}\label{subsec:freefree}

Inside the inner cavity, we detect a compact source at the position of the star at the four wavelengths. This source was previously detected at 0.87 mm and 1.3 mm \citep{and16,hua18a}, and we now detect it at 2.1 mm and 3.1 mm as well. As shown in Figure \ref{fig:highresprofiles}, the emission of this source increases significantly at longer wavelengths. At 3.1 mm, the brightness temperature at the disk center reaches values a factor of 2 higher than at the other bands. While these images are not directly comparable because of the different beam sizes,
the larger beam at 3.1 mm would in fact decrease the brightness temperature of a bright compact source due to beam smearing. Therefore, the increase in brightness temperature at longer wavelengths is likely not only real, but actually higher than seen in our observations.

In order to shed more light on the nature of this source, we can inspect the spectral index of the emission at this position. Using the spectral index maps from the multiband images, we estimate a spectral index of $\sim1.8$, $\sim1.3$, and $\sim0.6$ at 0.87-1.3 mm, 1.3-2.1 mm, and 2.1-3.1 mm, respectively. At 0.87-1.3 mm, the spectral index in fact remains constant up to $r\sim20$ au, and is hence likely associated with dust thermal emission. However, the significant flattening of the spectral index at 1.3-2.1 mm and 2.1-3.1 mm cannot be caused by dust emission, even if considering the effects of self-scattering \citep{zhu19}. Instead, these spectral slopes indicate that the emission at 2.1 mm and 3.1 mm has a significant contribution of free-free emission. We discuss the origin of this free-free emission in Sect. \ref{subsec:discfreefree}.

\section{Dust characterization}\label{sec:mod}

As mentioned above, multiwavelength observations of the dust thermal emission can be used to characterize the dust density, dust temperature, and dust particle size distribution \citep[e.g.,][]{per15,car19,mac19}. In order to analyze our observations, we aim at modeling the radial intensity profiles at each wavelength assuming an axisymmetric protoplanetary disk. The disk is also assumed to be geometrically thin and vertically isothermal, both good approximations when analyzing the (sub)millimeter emission of disks, which is in principle tracing the thin midplane layer of the disk where large dust grains have settled \citep{pin16}. The possible small errors that this approximation could introduce are minimized by the close to face-on geometry of the disk of TW Hya. Given these assumptions, the emergent intensity at a particular radius can be computed as a 1D vertically isothermal slab \citep{sie19,car19}:
\begin{equation}
    I_{\nu} = B_{\nu}(T_d) [(1-\textrm{exp}(-\tau_{\nu}/\mu))+\omega_{\nu}F(\tau_{\nu},\omega_{\nu})],
\end{equation}
where $T_d$ is the dust temperature, $B_{\nu}(T_d)$ is the black body emission at $T_d$ and frequency $\nu$, $\tau_{\nu}=\Sigma_d \chi_{\nu}$ is the optical depth, $\Sigma_d$ is the dust surface density, $\chi_{\nu}$ is the total dust opacity (i.e., absorption, $\kappa_{\nu}$, plus scattering, $\sigma_{\nu}$), $\omega_{\nu}=\sigma_{\nu}/(\kappa_{\nu}+\sigma_{\nu})$ is the dust albedo, and $\mu=cos(i)$ (where $i$ is the inclination angle, $i=0^{\circ}$ being face-on). In turn, $F(\tau_{\nu},\omega_{\nu})$ is defined as:
\begin{dmath}
    F(\tau_{\nu},\omega_{\nu})=\frac{1}{\textrm{exp}(-\sqrt{3}\epsilon_{\nu}\tau_{\nu})(\epsilon_{\nu}-1)-(\epsilon_{\nu}+1)}
    \times \left[\frac{1-\textrm{exp}(-(\sqrt{3}\epsilon_{\nu}+1/\mu)\tau_{\nu})}{\sqrt{3}\epsilon_{\nu}\mu+1} +
    \frac{\textrm{exp}(-\tau_{\nu}/\mu)-\textrm{exp}(-\sqrt{3}\epsilon_{\nu}\tau_{\nu})}{\sqrt{3}\epsilon_{\nu}\mu-1}\right] ,
\end{dmath}
where $\epsilon_{\nu}=\sqrt{1-\omega_{\nu}}$. Following \citet{car19}, we also include an approximation for anisotropic scattering by replacing the scattering coefficient $\sigma_{\nu}$ with an effective scattering coefficient defined as $\sigma^{\rm{eff}}_{\nu}=(1-g_{\nu})\sigma_{\nu}$, where $g_{\nu}$ is the forward scattering parameter.

The key to fitting this equation to multiwavelength observations is that a dust opacity law can be computed from a particular dust composition and particle size distribution \citep{bec90}. For a fixed composition, and assuming that the particle size distribution follows a power law with slope $p$ ($n(a)\propto a^{-p}$, where $a$ is the particle radius), the dust absorption and scattering opacity laws can be computed varying just the maximum particle size ($a_{max}$) and slope of the particle size distribution ($p$). \footnote{In principle one can also vary the minimum particle size, but the effect of this quantity in the (sub)millimeter opacity is minimal as long as it remains small. We fix this value to $0.01$ $\mu$m.} The value of $p$ is usually assumed to be $3.5$ \citep[e.g.,][]{mac18,car19,ued20}, since this is the value measured in the ISM \citep{mat77}, and is the expected value from a dust population produced by a collisional cascade \citep{doh69}. However, dust evolution models predict radial variations in this slope, with slightly flatter slopes at the position of the rings \citep[i.e., dust particle size distributions with more weight on the large particles;][]{dra19}. Therefore, 
we fit our observations using two different approaches: a model with the slope fixed to the standard $p=3.5$, as well as a model where we let $p$ be a free parameter \citep[e.g.,][]{mac19}.
Additionally, we adopt the dust composition used by the DSHARP program, computed with the Python package \texttt{dsharp_opac} \citep{bir18}. 

Following these standard assumptions, the emergent intensity at a particular radius and at any given wavelength in the (sub)millimeter can be hence computed as a function of just three or four parameters: the dust temperature $T_d$, the dust surface density $\Sigma_d$, the maximum grain size $a_{max}$, and the slope of the particle size distribution $p$. We adopt a Bayesian approach to obtain the posterior probability distributions of the model parameters at each radius. We use a standard log-normal likelihood function:
\begin{equation}
    \textrm{ln}P(\bar{I}(r)|\Theta)=-0.5\sum_i \left( \left( \frac{\bar{I_i}-I_{m,i}}{\hat{\sigma}_{\bar{I},i}} \right)^2 + \textrm{ln}(2\pi\hat{\sigma}_{\bar{I},i}^2)\right) ,
\end{equation}
where $\Theta$ is the vector of the model parameters ($T_d$, $\Sigma_d$, $a_{max}$, and $p$), $\bar{I_i}$ is the azimuthally averaged intensity at radius $r$ at frequency $\nu_i$, $I_{m,i}$ is the model intensity at this same radius and frequency, and assuming that the uncertainty $\hat{\sigma}_{\bar{I},i}$ at radius $r$ is:
\begin{equation}
    \hat{\sigma}_{\bar{I},i} = \sqrt{\sigma_{\bar{I},i}^2 + (\delta_i \bar{I}_i)^2},
\end{equation}
where $\sigma_{\bar{I},i}$ is the error of the mean obtained from the azimuthally averaged intensity profiles (see Sect. \ref{sec:results}), and $\delta_i$ is the absolute flux calibration error at each frequency. We set this systematic error to the nominal values of $10\%$ at Band 7, and $5\%$ at Bands 6, 4, and 3. 

Since we are fitting different observations, it is crucial to ensure that differences in angular resolution between the observations are minimized. To this end, new images with a common beam size of $0\rlap.''05\times0\rlap.''05$ were produced. This was done by first exploring different robust and uvtaper parameters that yielded beam sizes slightly smaller than $0\rlap.''05\times0\rlap.''05$, and then using the \texttt{CASA} task \texttt{imsmooth} to convolve these images to the desired common beam. Our model is fitted to the azimuthally averaged intensity profiles from these images, starting at 5 au to avoid the inner regions with free-free contribution. We note that we do not convolve the model, since each radius is fitted independently. Therefore, beam dilution might be affecting our results, although at relatively small spatial scales given the high spatial resolution of our observations. 

Given the low number of free parameters, we can compute the posteriors by simply exploring the parameter space with a uniform grid within reasonable values. We vary $T_d$, $\Sigma_d$, and $a_{max}$ in a $115\times225\times100$ grid, respectively. The model with a varying $p$ additionally explores 60 values of this parameter. $T_d$ is varied from $4$ to $50$ K, in steps of $0.4$ K. $\Sigma_d$ and $a_{max}$ are explored in logarithmic space in steps of $0.02$ dex from $10^{-3}$ g cm$^{-2}$ to $10^{1.5}$ g cm$^{-2}$, and in steps of $0.05$ dex from $10^{-3}$ cm ($10~\mu$m) to $100$ cm, respectively. Finally, $p$ is varied from 2.5 to 4.3 in steps of 0.03.

As mentioned above, our model is fitted to each radius independently, without imposing a particular temperature or surface density profile.
In order to ensure that there is a reasonable thermal balance between the disk temperature structure and the expected stellar irradiation (i.e., that the temperature in the disk does not reach physically unrealistic values), 
we use a prior for the dust temperature at each radius based on the expected temperature profile of a passively irradiated flared disk in radiative equilibrium \citep{chi97,dul01}:
\begin{equation}\label{eq:temp_prof}
    T_d(r)= \left( \frac{\varphi L_{\star}}{8\pi r^2 \sigma_{SB}} \right)^{0.25} , \\
\end{equation}
where $\varphi$ is the flaring angle of the disk, $L_{\star}$ is the stellar luminosity, and $\sigma_{SB}$ is the Steffan-Boltzmann constant. 
We compute this prior by varying $\varphi$ within the typical values $0.01-0.06$ \citep[e.g.,][]{hua18b}, and assuming a normal distribution for $L_{\star}$ centered at $0.23$ L$_{\odot}$ \citep{sok18} with a standard deviation of $15\%$ (i.e., $\sim0.035$ L$_{\odot}$).
We note that an alternative approach could be to fix the temperature at each radius to the value predicted by this equation \citep[e.g.,][]{lon20}, but 
letting the temperature be a free parameter in the model allows us to introduce the inherent uncertainty of eq. \ref{eq:temp_prof} into the posteriors of $\Sigma_d$, $a_{max}$, and $p$. 
The priors for the other three model parameters are set as uniform priors.

\begin{figure*}
\centering
\includegraphics[width=\textwidth]{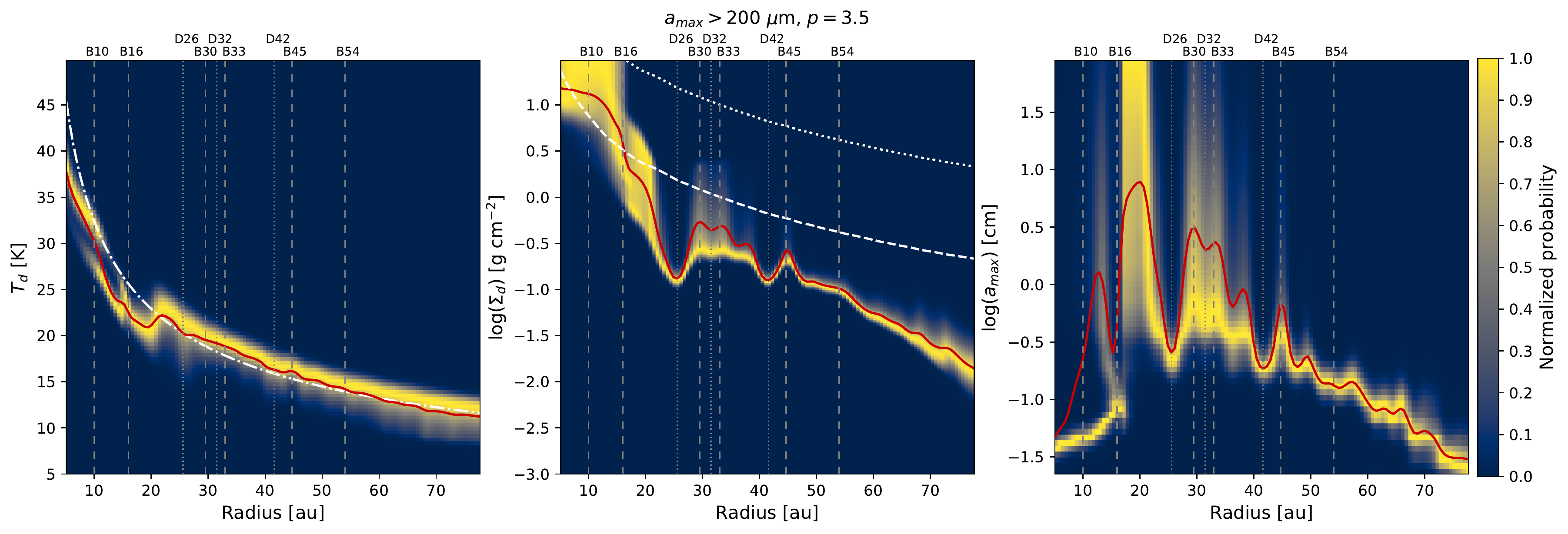}
\caption{Marginalized posterior probability distributions of the model parameters fitted to the multiwavelength observations assuming $a_{max}>200~\mu$m and with $p=3.5$. In order to facilitate its visualization, the posterior probabilities are normalized to the maximum marginalized probability at each radius. 
The vertical dashed and dotted lines indicate the positions of the bright and dark rings in the disk, respectively. The white dashed line in the left plot shows the temperature profile derived from eq.\ref{eq:temp_prof}, used as the reference for the temperature prior. The white dashed and dotted curves in the surface density plot show the dust surface density profiles for which the Toomre $Q$ parameter equals 1, assuming a gas to dust ratio of 100 or 10, respectively. The red curve in each panel shows the expected value of the corresponding parameters at each radius, computed using eq. \ref{eq:expvalue}.}
\label{fig:modlargedust_p3p5}%
\end{figure*}

\begin{figure*}
\centering
\includegraphics[width=0.8\textwidth]{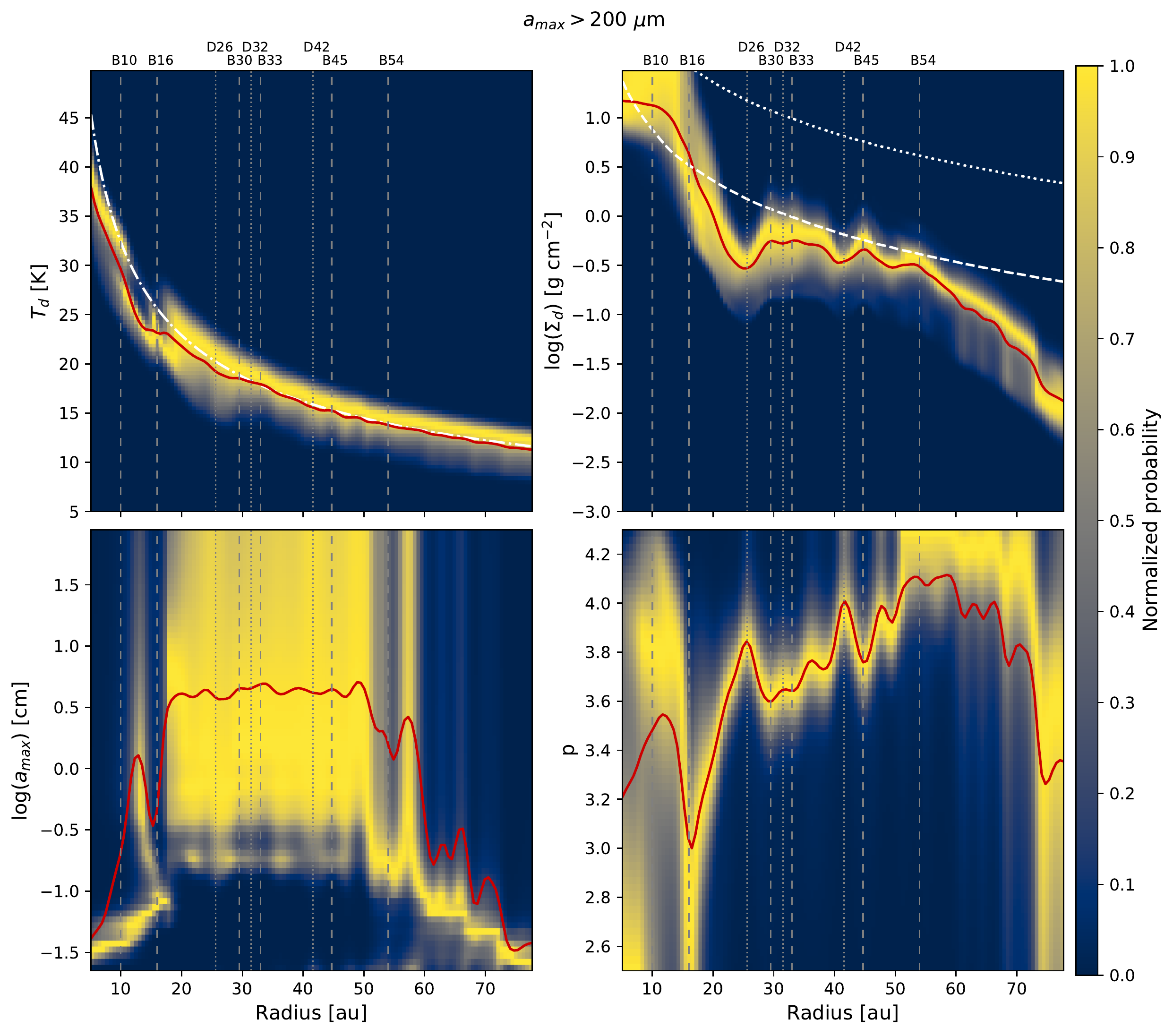}
\caption{Same as Figure \ref{fig:modlargedust_p3p5}, but for the model with $p$ as a free parameter.}
\label{fig:modlargedust}%
\end{figure*}

\subsection{Modeling results}

\begin{figure*}
\centering
\includegraphics[width=0.9\textwidth]{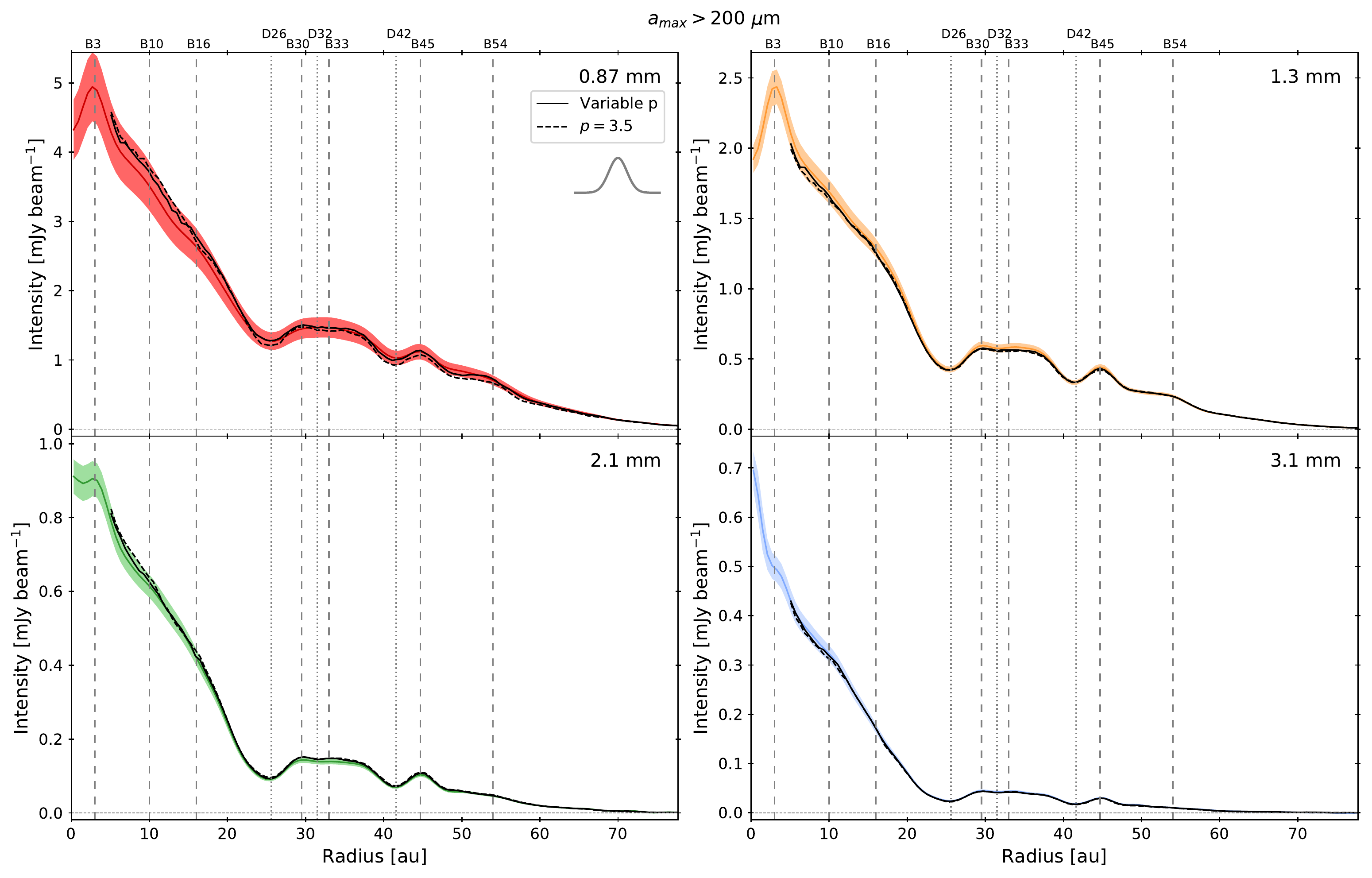}
\caption{Radial intensity profiles of the dust emission of the disk around TW Hya at 0.87 mm (top-left), 1.3 mm (top-right), 2.1 mm (bottom-left), and 3.1 mm (bottom-right) convolved to a circular $0\rlap.''05$ beam. The colored shaded regions indicate the error bars at each radius (including the flux calibration error). The solid and dashed lines show the model intensity profile with maximum posterior probability for $a_{max}>200~\mu$m, using a varying $p$ or fixing it to $p=3.5$, respectively. The Gaussian inset in the top-left panel displays the beam shape (the same at the four wavelengths). Our model is able to accurately reproduce the observed profiles at all radii except $r<20$ au.}
\label{fig:maxlikeprofileb}%
\end{figure*}

The first important result to note is that, at most radii, the radial intensity profiles can be relatively well reproduced with two different families of solutions: one solution with $a_{max}>1$ mm throughout the disk, and another one with $a_{max}$ ranging from $\sim10~\mu$m to $\sim50~\mu$m. The former solutions present a higher (sub)millimeter opacity, so they predict lower $\Sigma_d$ and higher $T_d$. On the other hand, $a_{max}$ between $10~\mu$m and $50~\mu$m require significantly higher $\Sigma_d$ and lower $T_d$. In order to avoid the blurring of the posteriors by these two families of solutions, we compute the posterior probability distributions after splitting the $a_{max}$ range into small ($10~\mu$m$<a_{max}<200~\mu$m) and large particles ($200~\mu$m$<a_{max}<10$ cm). The resulting marginalized posterior probability distributions of $T_d$, $\Sigma_d$, $a_{max}$, and $p$ at each radius are shown in Figures \ref{fig:modlargedust_p3p5} and \ref{fig:modlargedust} for $a_{max}>200~\mu$m, and Figures \ref{fig:modsmalldust_p3p5} and \ref{fig:modsmalldust} for $a_{max}<200~\mu$m. The red curve in these plots indicates the expected value of each parameter, computed as:
\begin{equation}\label{eq:expvalue}
    E(X) = \frac{\sum_i X_i P(X_i|\bar{I}(r))}{\sum_i P(X_i|\bar{I}(r))},
\end{equation}
where $X_i$ is each value of the parameter within our grid, and $P(X_i|\bar{I}(r))$ is the marginalized posterior probability of the parameter within each cell of our grid. For comparison, we also plot the dust surface density profiles for which the Toomre $Q$ parameter would be equal to 1, assuming two different values of the gas-to-dust mass ratio (100 and 10). We compute the Toomre $Q$ parameter as:
\begin{equation}\label{eq:toomre}
    Q = \frac{c_s \Omega_K}{\pi G \Sigma_g},
\end{equation}
where $c_s=\sqrt{k_B T / \mu m_p}$ is the sound speed, $\Omega_K=\sqrt{G M_{\star} / r^3}$ is the Keplerian frequency, and $\Sigma_g$ is the gas surface density. We use the expected value of the temperature obtained from eq. \ref{eq:expvalue}, as well as a stellar mass of $0.6~M_{\odot}$ \citep{sok18}. A disk with a dust surface density above these plotted curves would likely be gravitationally unstable and would show nonaxisymmetric features such as spiral arms. The high degree of axisymmetry of the disk of TW Hya indicates that its density profile should be below these limits. In addition, Figures \ref{fig:maxlikeprofileb} and \ref{fig:maxlikeprofiles} show the observed radial intensity profiles (convolved to the common $0\rlap.''05$ circular beam), as well as the intensity profile of the model with maximum posterior probability at each wavelength, for $a_{max}>200~\mu$m and $a_{max}<200~\mu$m respectively. 

As shown in these figures, our results with $a_{max}<200~\mu$m would require a significantly low gas-to-dust mass ratio ($\sim20$) for the disk to be gravitationally stable. Furthermore, our model predicts extremely low temperatures (see Sect. \ref{subsec:twofamilies}). Overall, the physical conditions predicted with $a_{max}=10-50~\mu$m appear to be physically unrealistic, so we favor our results with $a_{max}>200~\mu$m as more accurate estimates of the dust content of the disk of TW Hya. In the following we focus on this family of solutions, and in Sect. \ref{subsec:twofamilies} we explore how to further discern between these two ranges of $a_{max}$.

When inspecting the radial variation of our resulting posteriors, these can be separated into two different regions. At radii $>20$ au, our model successfully reproduces the observed intensity profiles (see Fig. \ref{fig:maxlikeprofileb}).
On the other hand, at $r\sim19$ au, the posteriors display a sharp jump between the physical properties of the inner disk and those of the outer disk. Furthermore, within this radius, our model is unable to reproduce the intensity profile at the four wavelengths. Interestingly, this region corresponds to the region that shows spectral indices$<2.0$ (see Fig. \ref{fig:spindex}), indicating that the optical depth is extremely high and dominated by self-scattering \citep{zhu19}. 

At $20<r<60$ au, our $p=3.5$ model (Fig. \ref{fig:modlargedust_p3p5}) robustly constrains $a_{max}$ at most radii, with values that range from $\sim1$ mm to $\sim10$ cm. The value of $a_{max}$ also displays a clear decreasing trend with radius, consistent with the effects of radial drift and/or the reduction of the fragmentation barrier with radius \citep{bra08}. 
At $r>60$ au, where only the 0.87 mm and 1.3 mm continuum emission are detected with good S/N, the model predicts a maximum grain size that decreases down to $\sim300~\mu$m. The model shows evidence of smaller grain sizes at the position of the gaps, with relatively significant contrast between rings and gaps.
We note, however, that the $p=3.5$ model appears to be unable to constrain $a_{max}$ within the $1-10$ cm range at the position of the rings B30 and B33, which translates into higher uncertainties in the dust surface density at these positions.

On the other hand, the model with $p$ as a free parameter displays significant differences at $20<r<60$ au. Interestingly, once the slope of the particle size distribution is allowed to vary, the model is unable to constrain $a_{max}$ within the $0.1-100$ cm range. Instead, the slope $p$ is robustly constrained, with a positive trend with radius (from $\sim3.5$ to $\sim4.1$), steeper slopes at the position of the gaps, and flatter ones at the rings. 
The implications from this model are in fact similar to the trend in $a_{max}$ shown by the $p=3.5$ model: the abundance of large particles decreases with radius and at the position of the gaps.

The dust surface density, $\Sigma_d$, is well constrained at $r>20$ au for both models. Its radial morphology displays a general decreasing trend with radius, with lower values at the gaps and higher at the rings. The $p=3.5$ model predicts slightly lower values of $\Sigma_d$ than the case with varying $p$. On the other hand, no clear trends are seen in the temperature profile, with its posterior being set mostly by the prior.
We note that most gaps in TW Hya are narrow, with D25 being the wider and better resolved one. Therefore, beam dilution might affect our results, particularly at the position of the gaps. Finally, in order to help visualize the trends and correlations between the different parameters, we also show in Figures \ref{fig:corner1}, \ref{fig:corner2}, and \ref{fig:corner3} three corner plots with the 2D posteriors at some representative radii (25 au, 30 au, and 45 au). 

\section{Discussion} \label{sec:discussion}

\subsection{Free-free emission inside the 3 au cavity}\label{subsec:discfreefree}

A compact source inside the 3 au inner cavity, at the position of the star, is detected at the four wavelengths. While this source might be associated with dust thermal emission at 0.87 and 1.3 mm, its spectral index at 1.3-2.1 mm ($\alpha\sim1.3$) and 2.1-3.1 mm ($\alpha\sim0.6$) indicates that the emission at these wavelengths is dominated by free-free emission. The centimeter-wavelength SED of TW Hya shows some evidence of free-free emission \citep{pas12}, although with an estimated relatively small contribution of $\sim0.1$ mJy ($\sim5\%$) at 9 mm according to \citet{men14}. This emission could be associated with gas from an ionized radio jet \citep[e.g.,][]{rey86}, from a photoionized wind \citep[e.g.,][]{pas12}, or from both \citep[e.g.,][]{mac16}. The elongation and orientation of the free-free emission is sometimes used to determine the presence of a radio jet \citep[e.g.,][]{rod14}, but the low inclination of TW Hya makes it unfeasible to resolve any elongation in our data.

\begin{figure}
\centering
\includegraphics[width=0.48\textwidth]{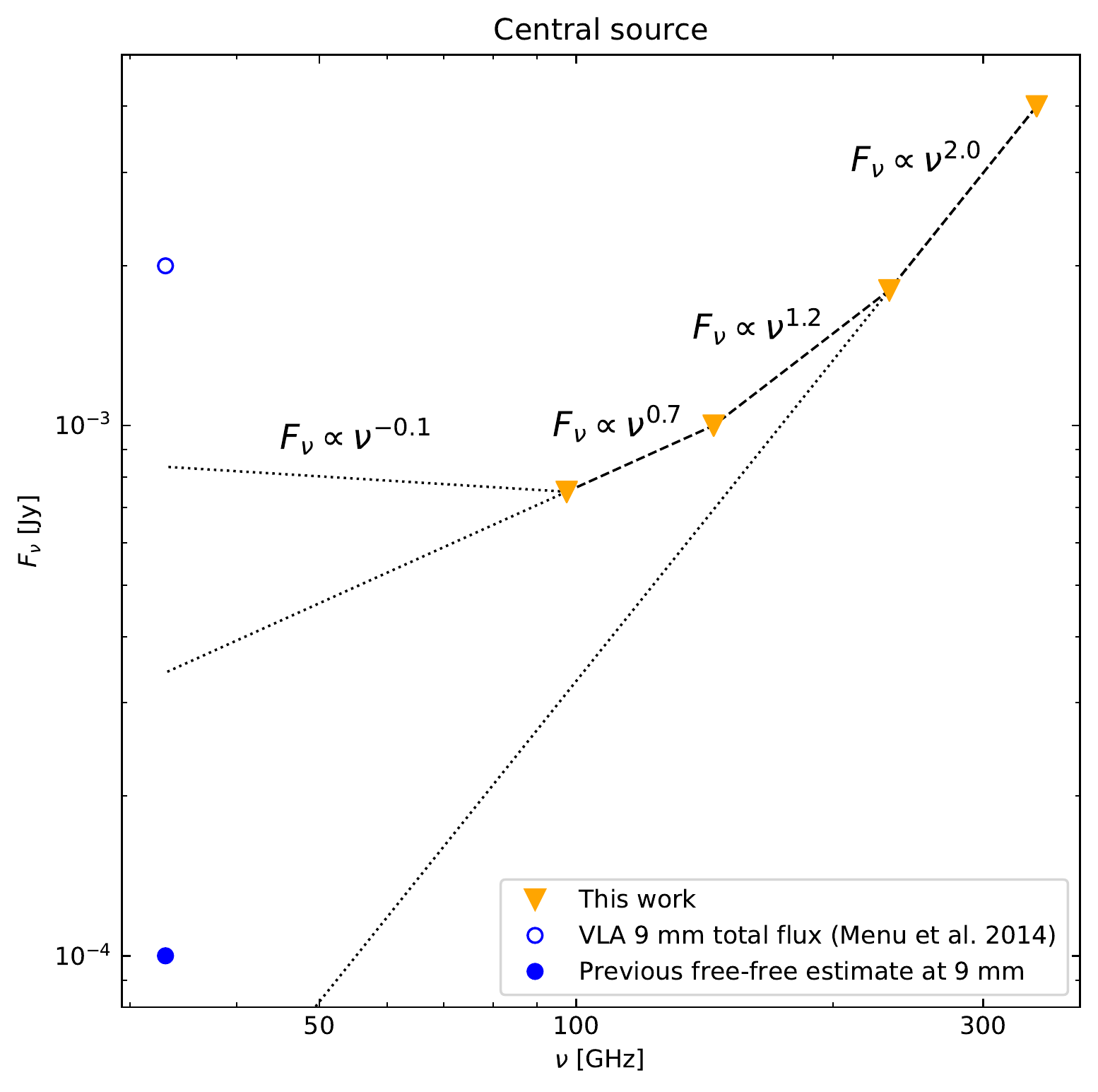}
\caption{Spectral energy distribution (SED) of the central source of emission located inside the 3 au cavity in TW Hya. The orange triangles indicate the fluxes estimated from our high angular resolution observations at 0.87 mm, 1.3 mm, 2.1 mm, and 3.1 mm (see insets in Fig. \ref{fig:maps}. The blue empty circle indicate the VLA 9 mm flux of TW Hya \citep{men14}, while the filled blue circle shows the estimated free-free contribution at this wavelength reported by these authors. The dashed lines depict the spectral indices of the central source inferred from our observations. The dotted lines show different extrapolations of this source's flux up to 9 mm, demonstrating that the free-free emission at this position is higher than previously thought.}
\label{fig:centralsource}%
\end{figure}

Using our high angular resolution images (see insets in Fig. \ref{fig:maps}), we can estimate the flux of the central source at 0.87 mm, 1.3 mm, 2.1 mm, and 3.1 mm. We do so by measuring the flux inside the $r<2$ au region around the star. These measurements are shown in Fig. \ref{fig:centralsource}. We note that our beam sizes are relatively similar to the size of the inner cavity ($r\sim3$ au), so
our flux estimates might include some dust contribution from the ring at 3 au. From these measurements we estimate a spectral index of $2.0\pm0.3$ at 0.87-1.3 mm, $1.24\pm0.16$ at 1.3-2.1 mm, and $0.72\pm0.18$ at 2.1-3.1 mm. These indices are very similar to the values found at the position of the star in the spectral index maps (1.8, 1.3, and 0.6, respectively).
Given the flat 2.1-3.1 mm spectral index, the dust contribution at these wavelengths is likely small. Using the 3.1 mm flux, and assuming a constant $2.1-9$ mm spectral index of $0.72\pm0.18$, we would obtain a 9 mm free-free contribution of $0.34\pm0.06$ mJy, a factor 3.4 higher than previously estimated. While centimeter emission is known to be variable \citep[e.g.,][]{uba17,esp19}, variability is unlikely to be the cause of this large discrepancy. Therefore, despite the moderate uncertainties in this estimate, our observations indicate that the free-free contribution at 9 mm was likely underestimated, and that it is closer to $\sim20\%$ of the total flux at this wavelength.

Interestingly, the emission from ionized jets has often been assumed to become optically thin at $\lambda\sim7-10$ mm, so its spectral index should become $\sim-0.1$ at shorter wavelengths \citep{rey86,ang18}. Additionally, the free-free emission from a photoionized wind is expected to be optically thin at millimeter and centimeter wavelengths, hence presenting a flat spectral index too \citep{pas12}. Such a flat spectral slope would imply a relatively constant flux from 3 mm to 9 mm. Extrapolating the previous 9 mm free-free estimate of $\sim0.1$ mJy to 3.1 mm would imply a very small contribution compared to the peak intensity at this position. This low emission would not be able to flatten the spectral index up to 0.6. Similarly, extrapolating the 3.1 mm flux to 9 mm using a flat spectral index would imply a free-free contribution $\sim40\%$ the total flux at this wavelength. Such a high contribution from a central compact source seems extremely unlikely given the visibility profile of the 9 mm data \citep{men14}. Therefore, our observations appear to indicate that the spectral index of the central component remains positive at least up to 3.1 mm, probably within the $0.4-0.8$ range. If confirmed, this would suggest that the free-free emission is arising mostly from an ionized jet which still presents an optically thick component at 3.1 mm. 

The contribution from radio jets at these short wavelengths has usually been neglected, mostly due to the fact that dust emission is expected to dominate at these wavelengths.
However, the high angular resolution of our observations, paired with the lack of dust emission in the inner 3 au of TW Hya, have allowed us to estimate the free-free contribution in TW Hya at millimeter wavelengths. A detailed analysis of this potential radio jet is out of the scope of this paper, but its presence might have interesting implications. Assuming a conical jet, the turnover frequency at which the free-free emission transitions from partially optically thick to completely optically thin is correlated with the injection radius of the jet \citep{rey86}. The turnover frequency implied by our results is higher than usually assumed \citep[e.g.,][]{ang18}, and it would imply a smaller injection radius than normally considered. Higher angular observations at longer wavelengths are needed to confirm and further analyze the origin of the free-free emission in TW Hya.

\subsection{Dust grain size in TW Hya: small ($a_{max}<50~\mu$m) or large ($a_{max}>1$ mm)}\label{subsec:twofamilies}

\begin{figure*}
\centering
\includegraphics[width=\textwidth]{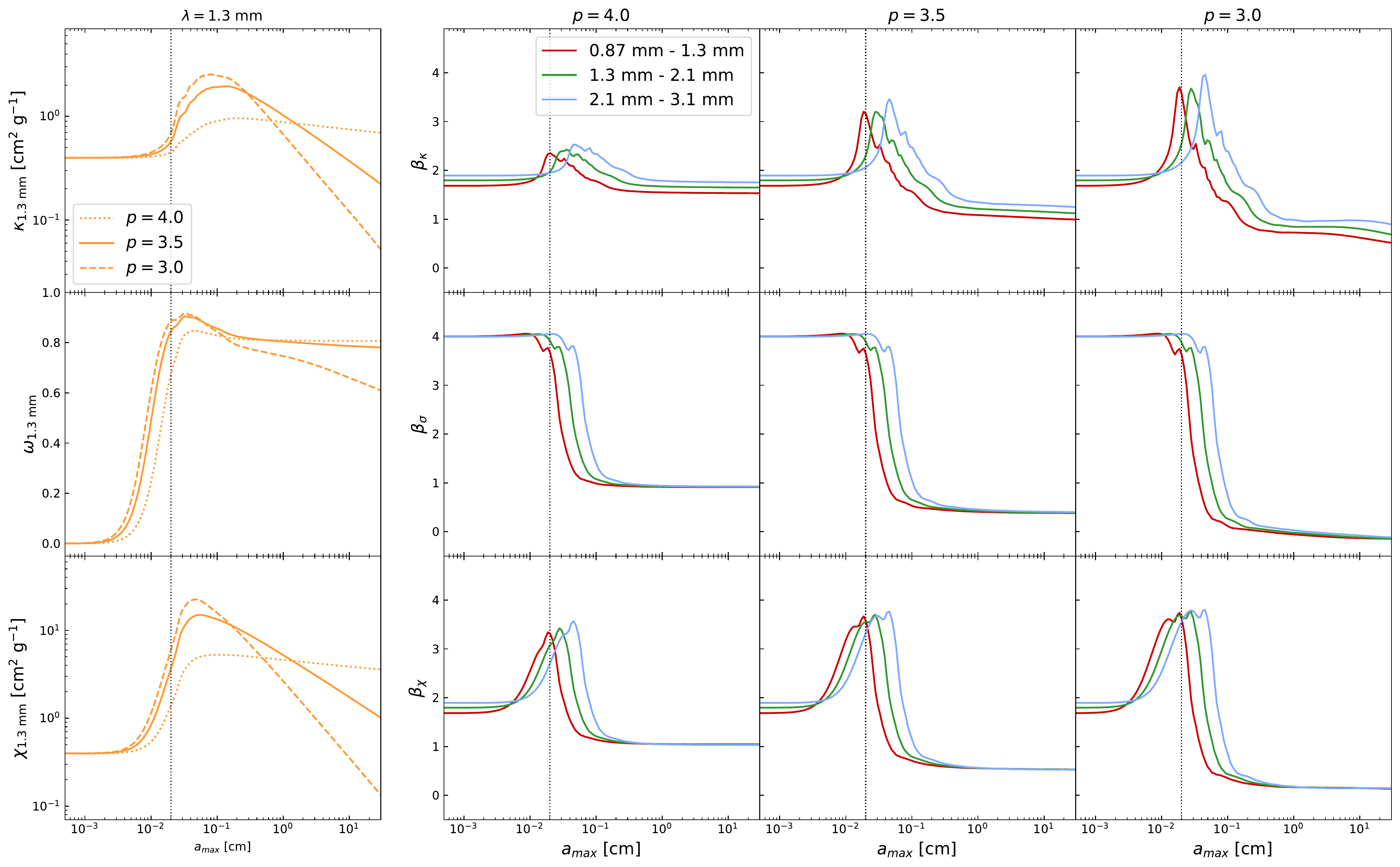}
\caption{Opacity laws used in the modeling of the multiwavelength observations. The left panels show the absorption opacity (top), albedo (middle), and total extinction (bottom) at 1.3 mm as a function of maximum grain size ($a_{max}$), for three values of the power-law index of the particle size distribution ($p$). The nine panels on the right show the spectral index of the absorption opacity ($\beta_{\kappa}$; top), the scattering opacity ($\beta_{\sigma}$; middle), and the total extinction ($\beta_{\chi}$; bottom) between the different wavelengths of our data, as a function of $a_{max}$, and for $p=4.0$ (left column), $p=3.5$ (middle column), and $p=3.0$ (right column). The vertical dashed line in all the panels indicates $a_{max}=200~\mu$m, the limit we set to separate between the two families of solutions in our analysis. }
\label{fig:beta}%
\end{figure*}

Our model can in principle reproduce the data with $a_{max}=0.1-10$ cm or $a_{max}=10-50~\mu$m. The appearance of these two families of solutions is caused by the frequency dependency of the dust opacity. In order to show this, we plot in Figure \ref{fig:beta} the dust opacity law used in our analysis \citep{bir18}. This figure shows the 1.3 mm absorption opacity ($\kappa$), albedo ($\omega$), and total extinction ($\chi$) as a function of maximum grain size for three different values of $p$, as well as the spectral index ($\beta$) of the absorption opacity, scattering opacity ($\sigma$), and total extinction between the different wavelengths of our data (i.e., assuming $\kappa\propto\nu^{\beta_{\kappa}}$, $\sigma\propto\nu^{\beta_{\sigma}}$, and $\chi\propto\nu^{\beta_{\chi}}$). The dust opacity law might not necessarily follow a power-law (as suggested by the fact that $\beta$ varies significantly with wavelength), but computing its spectral index provides a good visualization of its trend with frequency. 
As can be seen in this figure, for $a_{max}<50~\mu$m and for $a_{max}>1$ mm,  $\beta_{\chi}$ remains relatively constant with wavelength between 0.87 mm and 3.1 mm. In other words, our observations do not appear to show evidence for strong changes of $\beta_{\chi}$ within the covered frequency range, so our data can in principle be reproduced with values of $a_{max}$ in these two ranges.

The actual values of $\chi$ and $\beta_{\chi}$ in these two $a_{max}$ ranges are substantially different nonetheless. Therefore, the predicted values of $\Sigma_d$ and $T_d$ required to reproduce the data differ significantly. The total extinction for $a_{max}=10-50~\mu$m is more than one order of magnitude lower than for $a_{max}>1$ mm, and $\beta_{\chi}$ also differs from $\sim1.8$ to $\sim0.7$, respectively. As a consequence, the models with $a_{max}=10-50~\mu$m require extremely high values of $\Sigma_d$, higher than 1 g cm$^{-2}$ for radii $<55$ au. As shown in Figure \ref{fig:modsmalldust}, if a standard gas-to-dust mass ratio of 100 is assumed, these high dust surface densities would result in a gravitationally unstable disk up to $\sim60$ au (i.e., Toomre $Q$ parameter $>1$). 
If this was the case, we would expect the disk to display nonaxisymmetric features. 
The high degree of axisymmetry of TW Hya, together with the current estimates of its gas surface density \citep[e.g.,][]{zha17}, imply that local dust-to-gas mass ratios $\gtrsim10$ would be required to explain the high dust surface densities predicted with $a_{max}=10-50~\mu$m.
Additionally, the solutions with $a_{max}=10-50~\mu$m require an extremely cold disk, with $T_d<15$ K at $r=20$ au. In contrast, midplane temperature estimations based on C$^{18}$O observations yield values around 27 K at 20 au \citep{zha17}, more in line with the temperature predicted with $a_{max}>200~\mu$m. The physical conditions obtained with $a_{max}=10-50~\mu$m are, thus, likely physically unrealistic, so in the remainder of this discussion we focus on the solutions with $a_{max}>200~\mu$m. 

\begin{figure}
\centering
\includegraphics[width=0.48\textwidth]{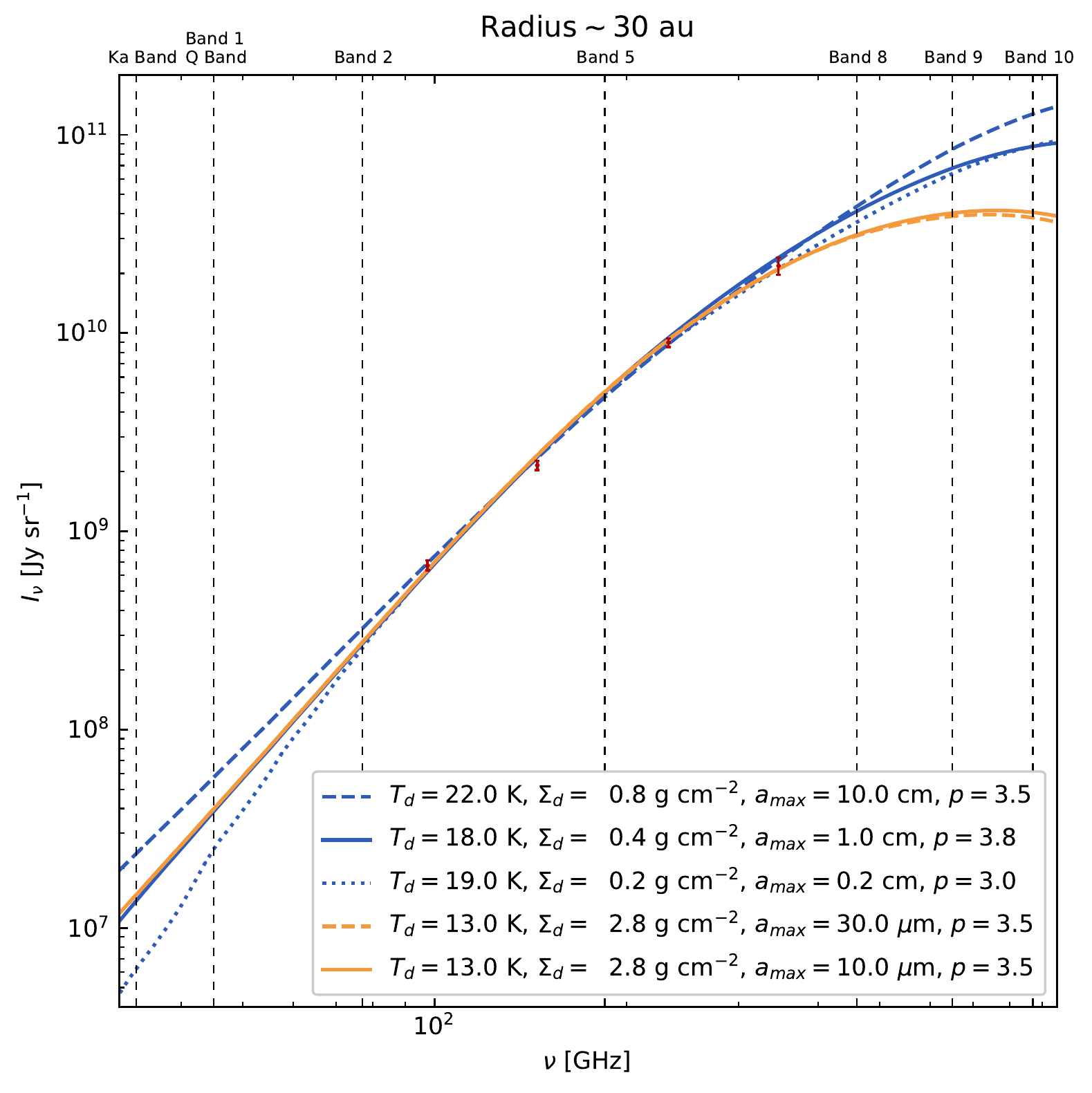}
\caption{Spectral energy distribution (SED) of the dust thermal emission at 30 au. The red points with error bars indicate the emission at 0.87 mm, 1.3 mm, 2.1 mm, and 3.1 mm at this radius. The different curves indicate the predicted SED for different combinations of parameters that are able to reproduce our data. The vertical dashed lines indicate the nominal frequencies of the ALMA Bands not used in this study, as well as the VLA Bands at $\lambda<1$ cm. As can be seen, observations at $\nu>400$ GHz (ALMA Bands 8 or higher), or at $\nu<50$ GHz (ALMA Bands 1 or 2, or VLA's Q Band and Ka Band) would help further constrain the particle size distribution in the disk.}
\label{fig:SED}%
\end{figure}

Nevertheless, these two families of solutions should be easily distinguishable with more observations. As shown in Figure \ref{fig:SED}, the emission predicted within these two $a_{max}$ ranges starts to differ significantly at frequencies higher than $\sim400$ GHz ($\lambda<0.75$ mm; ALMA Band 8; although see Sect. \ref{subsec:innerdisk}) because of the required low temperatures for $a_{max}=10-50~\mu$m. Furthermore, the emission differs slightly at frequencies lower than $\sim50$ GHz ($\lambda>6$ mm; ALMA Band 1, or VLA's Q Band and Ka Band) due to the steeper $\beta_{\chi}$ at this $a_{max}$ range. Alternatively, one could take advantage of the fact that both ranges of $a_{max}$ present very different values of the albedo at (sub)millimeter wavelengths: $\omega\sim0$ for $a_{max}=10-100~\mu$m, whereas $\omega\sim0.9$ for $a_{max}>200~\mu$m. These different albedos would produce distinctive continuum polarization patterns. In particular, if $a_{max}>1$ mm, the high albedos of these larger particles would produce linearly polarized emission dominated by self-scattering \citep{kat15}. So far, only a low S/N detection of (sub)millimeter polarized continuum emission toward the disk of TW Hya has been reported \citep{vle19}. The polarization angle of the emission shows a toroidal geometry, indicating an origin different than self-scattering. However, polarized emission is only detected in the outer regions of the disk ($r>40$ au). More observations of the continuum polarized emission of TW Hya at multiple wavelengths are needed to place further constraints on the dust maximum grain size based on dust self-scattering.

\subsection{The optically thick inner disk ($r<20$ au)} \label{subsec:innerdisk}

Our analysis predicts significantly different dust densities and maximum particle sizes in the innermost regions compared to the rest of the disk. The range of $a_{max}$ values predicted by our model at $r<20$ au jumps from $\gtrsim3$ mm to $\sim300-700~\mu$m, while $\Sigma_d$ reaches extremely high values of around $10$ g cm$^{-2}$. These values are entirely consistent with the results obtained by \citet{ued20}, who modeled the SED of the inner disk of TW Hya and found $a_{max}=300~\mu$m and $\Sigma_d=10$ g cm$^{-2}$ at 10 au. The ALMA observations modeled by these authors covered the same wavelengths analyzed in our study with the addition of 0.45 mm (ALMA Band 9), albeit at much lower angular resolution. Due to this lower resolution, their modeling focused only on the SED of the inner 10 au of the disk.

Despite the agreement between the analysis by \citet{ued20} and our results at $r<20$ au, the sudden decrease in particle size and the significant increase in dust surface density at $r<20$ au are hard to explain. Additionally, if the required high dust surface densities were real, the disk would probably become gravitationally unstable between $\sim5$ au and $\sim15$ au unless the gas-to-dust ratio at each radius was $\sim10$ (see Fig. \ref{fig:modlargedust}). In fact, as shown in Figure \ref{fig:maxlikeprofileb}, these inner regions are not as well reproduced by our model. 
The reason why our relatively simple model faces these issues in this region is likely related to the extreme optical depths at these radii. Our model assumes that the dust is located in a geometrically thin layer in the midplane, but in reality the dust presents a vertical density distribution. Dust settling results in the larger particles accumulating in the midplane, while smaller particles are present up to higher regions above it \citep[e.g.,][]{fro09}. If the disk is sufficiently dense, the vertical optical depth might reach values $>>1$ above the midplane, where larger particles are not present. In principle, one could still find a good fit to the data as long as the emission at the multiple wavelengths was tracing the same regions of the disk. However, as one moves to longer wavelengths the optical depth decreases. Thus, the emission traces regions slightly closer to the midplane, hence with larger dust grains, potentially different temperatures, and tracing different column densities. This effect was explored by \citet{sie20}, who showed that the combination of settling and high optical depth might indeed result in an incorrect estimate of the maximum grain size if $\Sigma_d\gtrsim3$ g cm$^{-2}$. In particular, these authors demonstrated that, because of dust settling, the emission of a protoplanetary disk with $a_{max}=1$ cm (and $p=3.5$) would present spectral indices $<2$ between 0.87 mm and 3 mm at radii where the dust surface density is high (i.e., radii with high optical depths). Estimating a value of $a_{max}$ from this emission without taking the dust vertical distribution into account would yield $a_{max}$ between $\sim300~\mu$m and $\sim600~\mu$m (see Fig. 8 in \citealp{sie20}). This range of $a_{max}$ coincides with the values obtained by our model at $r<20$ au and is also consistent with the results by \citet{ued20}, who did not include dust settling in their model.

\begin{figure}
\centering
\includegraphics[width=0.48\textwidth]{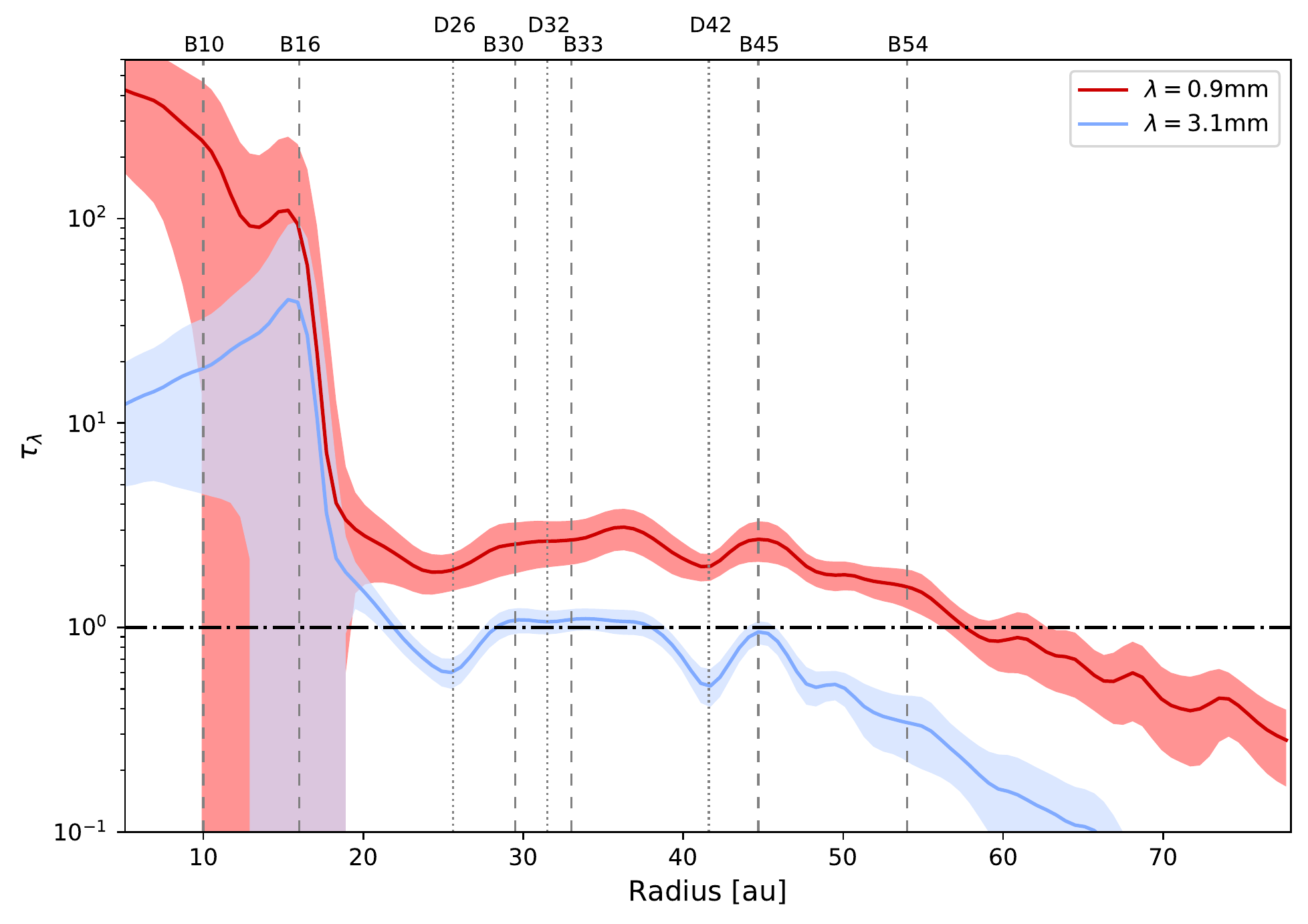}
\caption{Radial profile of the expected value of the optical depth at 0.87 mm (red) and 3.1 mm (blue), for our model with $a_{max}>200~\mu$m and a varying $p$. The expected value is computed from the posterior probability distribution of our model using eq. \ref{eq:tau}.}
\label{fig:tau}%
\end{figure}

In order to show whether this scenario is consistent with our observations of TW Hya, 
we computed the expected value of the optical depth at each radius as:
\begin{equation}\label{eq:tau}
    E(\tau_{\nu}) = \frac{\sum_{i} \chi_i(a_{max,i},p_i)~\Sigma_{d,i}~P(\Sigma_{d,i},a_{max,i},p_i|\bar{I}(r))}{\sum_i P(\Sigma_{d,i},a_{max,i},p_i|\bar{I}(r))} ,
\end{equation}
where $\chi_i(a_{max,i},p_i)$ is the dust extinction for each value of $a_{max}$ and $p$ in our grid, and $P(\Sigma_{d,i},a_{max,i},p_i|\bar{I}(r))$ is the posterior probability of our model at each cell of the grid, marginalized over $T_d$.
The results at 0.87 mm and 3.1 mm are shown in Figure \ref{fig:tau}. We note that the values of the optical depth at $r\lesssim19$ au are highly uncertain due to the poor quality of our fit in these regions. Nevertheless, it can be clearly seen that the jump at $r\sim20$ au is coincident with the region of the disk where the optical depth is larger than 1 at all wavelengths. Furthermore, the 0.87 mm optical depth at $r\sim20$ au starts to increase significantly, reaching values $>>100$. In other words, our observations are likely not probing the midplane at $r<20$ au, but a layer above it. 

At the same time, the potential change with wavelength in the height traced by our data can also be seen in the spectral index profiles. As shown in Fig. \ref{fig:spindex}, the region with $\alpha<2.0$ starts at lower radii when moving toward longer wavelengths. At $r<10$ au the spectral index is below 2.0 at all wavelengths, suggesting extremely high optical depths at the four bands. Between 10 and 20 au, our observations likely trace layers well above the midplane at 0.87 mm and 1.3 mm, but deeper layers at 2.1 mm and 3.1 mm. This change in wavelength might be the cause why $a_{max}$ appears to decrease toward smaller radii (see Fig. \ref{fig:modlargedust}). 

Despite the apparent consistency between our results and the high optical depth scenario, we cannot firmly conclude that the estimated $a_{max}$ values of $\sim300~\mu$m at $r<20$ au are wrong. Nevertheless, reconciling the dust densities and particles sizes of the inner and outer disk would require invoking a complex scenario. For example, the dust grains might experience a sudden and significant increase in fragmentation efficiency at $r\sim20$ au, resulting in a substantial decrease of maximum particle size paired with an increase in dust surface density because of the slower radial drift of these smaller particles. It is worth noting that \citet{pin17b} predicted a similar step in particle size and dust surface density when crossing the water snowline (assuming that ice-coated dust grains have a lower fragmentation efficiency). However, the temperature of the water snowline is $\sim180$ K, whereas the temperature in TW Hya at $r\sim20$ au should be $\sim22$ K. This temperature is tantalizingly close to the temperature of the CO snowline \citep{zha15}, but this volatile likely has a lower effect on dust coagulation and fragmentation than water given its lower abundance. Overall, we conclude that the high optical depth scenario is a more likely explanation of our results at $r<20$ au.

If confirmed, this effect will be a crucial impediment to the measurement of dust masses and particle size distributions in the inner regions of protoplanetary disks. Importantly, these are the regions where most planets are thought to be formed, including the planets in our own Solar System \citep{ray20}. Improving our understanding of planetesimal formation in these regions will depend on our ability to pierce into the dense and optically thick inner disk. Observing at longer wavelengths appears as the immediate alternative, but we note that if the 3 mm optical depth reaches values $>1$ at $r\sim20$ au, then the regions within 10 au might still be optically thick up to 7 mm. Observations at wavelengths between 7 mm and 1 cm are probably needed to better characterize the dust content in the inner regions of protoplanetary disks. 

Admittedly, TW Hya is particularly bright for its age, but it is not an outlier in terms of brightness when compared to younger and more distant disks. In fact, the inner regions of protoplanetary disks appear to have a similar surface brightness at 0.87-1.3 mm, with the disk total flux being set mostly by its radial extent \citep[e.g.,][]{tri17,and18b,fac19}. It is still possible that the inner regions of fainter disks are less optically thick, in which case observations of these systems might be better suited to characterize the dust content of the planet forming regions in disks. In any case, observations at longer wavelengths are crucial to improve the chances of piercing through the high optical depths of these regions. Right now, the Very Large Array (VLA) is the only observatory capable of providing reasonable sensitivity and angular resolution at these wavelengths, but these can only be achieved for a small number of bright disks. The extreme improvement in sensitivity and resolution expected from the Next Generation Very Large Array will be crucial to perform an accurate characterization of the dust content in the inner regions of protoplanetary disks in a larger and more representative sample \citep{and18c}. 

\subsection{The dust content at $r>20$ au and its impact on planet formation}

While the high optical depth of the inner disk prevents us from 
reproducing the multiwavelength data in those regions, our model is much more successful at $r>20$ au. 
Our results constrain fairly well the dust surface density, although the obtained maximum grain size differs significantly between the model with a fixed slope for the particle size distribution ($p=3.5$) and the model with a varying $p$.
As noted above, the $p=3.5$ model is able to constrain $a_{max}$ at almost all radii $>20$ au, with smaller grain sizes at the gaps, and larger ones at the rings (Fig. \ref{fig:modlargedust_p3p5}). On the other hand, if $p$ is allowed to vary the model can only set a lower limit of $\sim1$ mm to $a_{max}$. This model, however, is able to robustly constrain $p$ at $20<r<60$ au, with steeper slopes (i.e., higher $p$) at the gaps, and flatter slopes (lower $p$) at the rings (Fig. \ref{fig:modlargedust}). These apparently different results are in fact indicating similar physical conditions, since a flatter slope of the particle size distribution will result in a higher abundance of large dust particles.

The fact that a model with a varying $p$ is only able to set a lower limit to $a_{max}$ suggests that fixing $p$ could significantly affect the estimates of $a_{max}$ when analyzing multiwavelength data. Including $p$ as a free parameter allows us to account for the uncertainties on this parameter in the estimates of dust density and maximum grain size. Unfortunately, it will be challenging to further constrain the grain size even with more observations, since $a_{max}>1$ mm produce very similar SEDs at $\lambda<1$ cm (despite requiring dust surface densities that can be different by a factor of 2; see Fig. \ref{fig:SED}). Given the correlations between the four parameters in our model (see Figures \ref{fig:corner1}, \ref{fig:corner2}, and \ref{fig:corner1}), tighter constraints on the dust temperature would help narrow the posterior of $a_{max}$. These could be placed by using a more complete radiative transfer modeling that considers the irradiation of the disk by the central star. However, this analysis would require a significant computational effort that would probably limit the parameter space explored by the model.

Despite this uncertainty, our model places an important constraint: $a_{max}$ must be larger than 1 mm at $20<r<60$ au.
In recent years, multiwavelength and polarization observations of the (sub)millimeter dust emission have yielded conflicting results regarding the dust particle sizes in protoplanetary disks. While the spectral behavior of the dust emission indicates the presence of dust particles larger than 1 mm \citep[e.g.,][]{men14,per15,taz16,mac18}, the polarized emission of many disks has been interpreted as the result of self-scattering from particles that cannot be larger than $\sim100~\mu$m \citep{kat16a,kat16b,kat17,hul18,den19}. A possible explanation of this discrepancy could lie in the incorrect (yet common) assumption of optically thin emission made by some previous multiwavelength analysis. However, more recent studies considering the effects of optical depth recovered similarly large particles \citep{car19,mac19}.
Our analysis has considered the effects of self-scattering and optical depth, showing that dust particles must be larger than 1 mm in TW Hya. The cause of this discrepancy must hence lie on a more fundamental assumption. Recently, \citet{kir20} showed that the self-scattering polarization pattern of disks could be consistent with larger dust particles than previously estimated if nonspherical dust grains were assumed. A careful reanalysis of previous polarization observations is still required, but nonspherical grains appear as a promising scenario to reconcile the different dust particle measurements.

As mentioned above, our analysis also shows that the ringed substructures in TW Hya are reproduced with higher dust densities paired with higher abundances of large dust particles (denoted by the flatter slopes of the particle size distribution) at the position of the rings. 
We note that we have modeled images that are convolved to a circular $0\rlap.''05$ beam. Even though this resolution allows us to resolve the two widest gaps in TW Hya (D26 and D42, see Fig. \ref{fig:profiles}), some emission from the rings will contaminate these features. 
Therefore, the contrast in $\Sigma_d$ and in $a_{max}$ and/or $p$ between rings and gaps is probably larger than predicted by our model. Higher angular resolution observations are needed to better constrain the dust content in the gaps. 

Nonetheless, our results confirm that the rings in TW Hya are the result of accumulations of large dust particles. The radial variations in spectral index, with lower values at the rings and higher in the gaps, had been interpreted as a potential indication of this segregation of dust particles \citep{tsu16}. However, a similar trend could also be reproduced with variations in optical depth \citep{hua18a}. Our multiwavelength analysis resolves this dichotomy between larger particles and higher optical depth, providing new insights on the radial structure of the particle size distribution in the disk of TW Hya. 

The actual value of the maximum dust grain size is not only important to robustly constrain the optical depth and dust density in disks, but it also has important implications on the ability of disks to form planetesimals. Most current models of planetesimal formation predict that growth through the streaming instability is required to bypass the main barriers in dust coagulation, thus explaining the formation of planetary systems on reasonable timescales \citep[][and references therein]{ray20}. While alternative mechanisms such as secular gravitational instability \citep{you11} may also play a role in the formation of planetesimals, the streaming instability has the most success in reproducing many properties of the Solar System \citep{nes19,ray20}. The underlying physical mechanism behind this instability is the gas drag on dust grains (and its back-reaction onto the gas). Therefore, a concentration of millimeter- and centimeter-sized particles, which experience the highest gas drag forces, is required to trigger this instability \citep{car15,yan17}. Our results suggest that particles of this size are present in protoplanetary disks, and that rings substructures are ideal places to accumulate them and harbor the necessary conditions to trigger planetesimal formation. 

\subsection{Insights on the origin of disk substructures}

Our results can be used to constrain the origin of the ring substructures in TW Hya. Most physical mechanisms that have been proposed to form such substructures in protoplanetary disks involve the onset of a gas pressure bump trapping large dust particles (e.g., MHD effects: \citealp{joh09,bai14,bai17}; planet-disk interactions: \citealp{zhu14,bae17,zha18}). However, ring substructures in the (sub)millimeter emission of disks have also been proposed to form due to changes in the grain properties across different snowlines of volatiles \citep[e.g.,][]{oku16,pin17b}.
The exact impact of each particular snowline is still unclear, since different ices could have different sticking properties \citep[e.g.,][]{pin17b} and different physical processes could take place near the snowlines \citep[e.g.,][]{ros13,bit14}.

Given the potential uncertainties in the temperature profile predicted by our model, as well as the uncertainties in the sublimation temperature of volatiles, we cannot exactly pinpoint the position of the multiple snowlines in the disk. However, from the dust temperature profile predicted by our analysis, the only major volatile snowline that could coincide with the substructures at $r>20$ au is the CO one ($T_{sub}=23-28$ K; \citealp{zha15}). As mentioned in Sect. \ref{subsec:innerdisk}, we cannot firmly discard that the CO snowline has an effect in the change in particle size at $r<20$ au, but it would not be able to explain the rest of the disk substructures.
Furthermore, our analysis indicates that the ring substructures at $r>20$ au in TW Hya are associated with accumulations of larger particles, so we can at least discard effects that result in the formation of rings of smaller particles, such as the process dubbed as dust sintering \citep{oku16}. This process predicts that, as dust grains get close to a particular snowline, the sublimation of ices from their surface could make the grains more fragile, hence resulting in a more efficient fragmentation into smaller dust particles. Therefore, rings would be formed as local enhancements of dust density due to the accumulation of these smaller particles, which experience a weaker gas drag than their parent larger particles \citep{oku16}. Besides the jump in maximum particle size at $r\sim20$ au (see Sect. \ref{subsec:innerdisk}), our results are inconsistent with this scenario, as also suggested by other recent studies \citep[e.g.,][]{mac19,car19}. Therefore, we conclude that snowlines are unlikely to be the driving mechanism of most ring substructures in TW Hya, as already found in other protoplanetary disks \citep[e.g.,][]{hua18b,lon18,van19}.

Distinguishing between the different mechanisms that produce ring substructures associated with gas pressure bumps is more complicated \citep[e.g.,][]{rug16}. In principle, the contrast in dust surface density between rings and gaps could be used to place some constraints on the amplitude of the pressure perturbations, which could provide insights on its origin. Planet-disk interactions can produce a wide variety of gas density perturbations depending on the gas viscosity, the mass of the perturbing planet, or other disk properties \citep{zha18}. On the other hand, zonal flows driven by MRI turbulence should produce relatively shallow gas density perturbations ($\sim10-20\%$; \citealp{sim14}). The variation in disk resistivity caused by the transition in disk ionization fraction at the so-called dead zone might produce a more prominent gas pressure bump \citep{flo15}. However, this bump has also been predicted to result in the appearance of vortices \citep{rug16}.

Our relatively limited angular resolution prevents us from obtaining accurate estimates of this contrast in density, but we can estimate lower limits at the two wide gaps in TW Hya (D25 and D41).
From the predicted dust surface density, we estimate contrasts $\gtrsim2$. It is unclear how to translate from a contrast in dust surface density to the amplitude of the underlying gas density perturbation, but it seems unlikely that MRI-driven zonal flows are able to produce such deep gaps \citep{sim14}. The high degree of axisymmetry displayed by TW Hya also appears hard to reconcile with the dead zone scenario \citep{rug16}. This would leave planet-disk interactions as the most likely origin for the D25 and D41 gaps.

If confirmed, a planet origin for the ring substructures in TW Hya would in fact pose a somewhat circular conundrum: if dust traps are needed to form planets, and the ring substructures are caused by planets, then it is unclear how the first generation of planets formed. A general answer to this question requires the study of a larger sample of disks, which is hence out of the scope of this paper. Nonetheless, the older age of TW Hya might ease this tension if other physical mechanisms were the cause of disk substructures at an earlier evolutionary stage. These early substructures could have helped form the first generation of planets, whose interaction with the disk could have then dominated the formation of substructures in the system, shaping the disk into its current morphology.

In any case, we note that there still remain several uncertainties about the magnetic flux evolution or the different MHD effects in disks \citep{bai17}. For example, recent global MHD simulations including ambipolar diffusion predict the formation of ring substructures due to the onset of MHD winds, with contrasts that are much larger than for MRI-driven zonal flows \citep{rio20}. Measurements of the magnetic field's strength and morphology will be key to understand the importance of the many potential MHD effects. Although unresolved in our analysis, the remaining ring substructures between $\sim30$ and $\sim38$ au appear to be shallower and could potentially be associated with zonal flows. 
Higher angular resolution observations are needed to better characterize these substructures.

\subsection{On the total dust mass of TW Hya} \label{subsec:dustmass}

By integrating the expected value of the dust surface density obtained with our multiwavelength analysis, we can also obtain an estimate of the total mass of solids in the disk of TW Hya. As noted above, our results at $r<20$ au are likely incorrect, which prevents us from obtaining a robust total disk mass. However, we can measure the dust mass at $r>20$ au. From the posteriors for $a_{max}>200~\mu$m we obtain a dust mass at $r>20$ au of $158^{+17}_{-15}~M_{\oplus}$ for the model with $p$ as a free parameter, and $91^{+9}_{-8}~M_{\oplus}$ with a fixed $p=3.5$.

In order to obtain an estimate of the total dust mass we assume that the dust surface density profile at $r<20$ au follows a power law:
\begin{equation}
\Sigma_d (r) = \Sigma_{d,20} \left( \frac{r}{20~{\rm au}} \right)^{-\gamma},   
\end{equation}
where $\Sigma_{d,20}$ is the dust surface density at 20 au. By exploring values of $\gamma$ between 1 and 1.5, we estimate a range of total dust masses of $250-330~M_{\oplus}$ for the model with $p$ as a free parameter. The range of dust masses decreases to $210-310~M_{\oplus}$ if $p$ is fixed to 3.5.
In comparison, the total gas mass in TW Hya has been estimated to be around $0.05~M_{\odot}$ using \textit{Herschel} observations of HD line transitions \citep{ber13}. Our estimated dust masses for the model with a varying $p$ would thus translate into global gas-to-dust mass ratios of $50-70$, consistent with a slight depletion of gas due to MHD winds \citep[e.g.,][]{bai16} or photoevaporative winds \citep[e.g.,][]{owe11} at the advanced age of TW Hya. Following this same procedure we would estimate a range of dust masses of $880-1210~M_{\oplus}$ if $a_{max}<200~\mu$m, implying a gas-to-dust mass ratio $14-19$. As discussed above, these dust mass and gas-to-dust mass ratios are likely unrealistic (see Sect. \ref{subsec:twofamilies}). 
It should be noted that these estimates represent global gas-to-dust mass ratios, but much higher values are likely found locally. The radial extent of the millimeter dust ($\sim70$ au) is much lower than that of the gas ($\sim200$ au; \citealp{hua18a}), and dust settling is expected to significantly increase the dust abundance in the midplane compared to the disk atmosphere. Therefore, the local gas-to-dust mass ratio in the disk midplane at $r<50$ au could be much lower than $50$.

Demographic studies of protoplanetary disks performed with ALMA and earlier (sub)millimeter observatories suggest an apparent conflict between the measured dust masses of disks and the expected amount of solids in exoplanetary systems \citep[e.g.,][]{and05,naj14,man18}. In short, the dust mass of protoplanetary disks appears to be too low to form exoplanetary systems such as the ones detected so far. This discrepancy could indicate that planetesimals and planets form earlier than expected, and hence a large amount of solids in disks would already be accumulated in large bodies that are not traced by ALMA. This scenario would also be consistent with the prevalence of disk substructures in relatively young disks, assuming their origin is related to planet-disk interactions \citep[e.g.,][]{alm15,and18a}.

Nevertheless, an alternative explanation is that dust mass measurements from disk surveys are significantly underestimated. These measurements are obtained from a flux at (sub)millimeter wavelengths assuming that the dust emission is optically thin, a single dust temperature, and a single value for the dust opacity \citep[e.g.,][]{and13,ans16,pas16,bar16,cie19}. Recent studies modeling protoplanetary disk spectral energy distributions (SEDs) have shown that, if physical disk models are used, a factor $\sim3$ higher dust masses can be obtained \citep{bal19,rib20}. These models naturally included the effects of optical depth and more realistic dust temperature and density distributions, but still assumed a smooth disk structure as well as a constant maximum grain size with radius. The overall effect of disk substructures on such a modeling is unclear, but they could still potentially result in dust mass underestimates if  disk substructures are accumulating significant amounts of dust mass.

Even though our model is relatively simple (i.e., geometrically thin and vertically isothermal), our multiwavelength analysis does account for the presence of disk substructures, as well as a dust grain size distribution that varies with radius. Therefore, it is worthwhile comparing our dust mass estimates with the masses that one would obtain following the same procedure as (sub)millimeter disk surveys. Assuming the same standard dust opacity law ($\kappa=10$ cm$^2$ g$^{-1}$ at 1000 GHz, with $\beta=1$) and dust temperature (20 K) as \citet{and05}, a dust mass of $54\pm5~M_{\oplus}$ would be obtained from our 0.87 mm flux of $1540\pm150$ mJy. A similar dust mass of $57\pm6~M_{\oplus}$ would be obtained from our 1.3 mm flux of $575\pm30$ mJy. These masses are between a factor of 4.5 and 5.9 lower than the total dust mass estimated with our multiwavelength analysis. This difference is slightly higher than the differences found with SED modeling using physical disk models \citep{bal19,rib20}. Therefore, while accounting for optical depth effects and more accurate dust opacities (i.e., including absorption and self-scattering opacities) might be the biggest contributor to the underestimation of dust masses by the large (sub)millimeter surveys, accounting for radial changes in $a_{max}$ and $p$ might yield even higher dust masses. As shown in Figure \ref{fig:tau}, the optical depth at 0.87 mm is higher than 1 up to $r\sim55$ au, so the optically thin assumption is surely far from valid for TW Hya. 

Overall, our larger mass does alleviate the discrepancy between the mass of solids in the disk and the mass of exoplanetary systems, but the difference might still require the trapping of some solid mass in larger bodies. 
In any case, given the more advanced age of TW Hya and the potential planet origin for some of the substructures in its disk, it would not be surprising that a significant amount of solids is already in the form of planets and/or planetesimals.

\section{Summary and conclusions} \label{sec:conclusion}

We present new $\sim50$ milliarcsecond angular resolution ALMA observations of the 3.1 mm continuum emission of the protoplanetary disk around TW Hya. We pair the analysis of this new data with a reanalysis of archival high resolution ALMA data at 0.87 mm, 1.3 mm, and 2.1 mm. Our observations resolve the main substructures in the disk at the four wavelengths, including the central cavity at 3 au.

We find a bright compact source of free-free emission at the position of the star at 2.1 mm and 3.1 mm. The spectral index between 2.1 mm and 3.1 mm at this position is 0.6-0.7. Our estimated fluxes and spectral indices imply 
a higher free-free contribution at centimeter wavelengths than previously thought. The bright free-free emission at 3.1 mm also suggests that this source is likely associated with an ionized jet that still harbors an optically thick component at 3.1 mm. The presence of optically thick free-free emission at such a high frequency is usually neglected, especially for a protoplanetary disk with a relatively low mass accretion rate such as TW Hya. The free-free emission at 3.1 mm might suggest a smaller injection radius of its radio jet than usually assumed.

The multiwavelength high resolution observations have allowed us to characterize the dust content in the disk of TW Hya and within its substructures. Even though our data can in principle be reproduced with $a_{max}<50~\mu$m or $a_{max}>1$ mm, the former values for the particle size imply unrealistically high dust surface densities, as well as unphysically low dust temperatures. We hence conclude that the maximum grain size in TW Hya is $>1$ mm. Future observations at shorter or longer wavelengths could easily confirm this.

The inner 20 au of TW Hya appear to be too optically thick to accurately characterize its dust content. This region displays spectral indices $<2$, which likely indicate extremely high optical depths dominated by self-scattering. Due to the combined effect of this high optical depth and dust settling, our 1D slab model is unable to reproduce the data. Observations at $7-10$ mm are needed to study the optically thinner emission that can trace regions closer to the midplane. This effect might hinder our ability to study the most important planet-forming regions in protoplanetary disks.

At $r>20$ au, the maximum grain size at each radius is constrained to be $>1$ mm. Our model with the slope of the particle size distribution fixed to $p=3.5$ finds clear radial fluctuations in $a_{max}$ associated with the rings and gaps in the disk. When letting $p$ vary in our model, no radial trend in $a_{max}$ is seen. Instead, we find significant radial variations in $p$, with flatter slopes at the position of the rings. Overall, our analysis shows that the rings in TW Hya are produced by accumulations of millimeter- and centimeter-sized particles, likely associated with annular gas pressure bumps in the disk. These rings are therefore ideal locations to trigger the formation of larger planetesimals through the streaming instability.
    
Finally, we estimate a total dust mass $250-330~M_{\oplus}$, which would imply a global gas-to-dust mass ratio between 50 and 70. This dust mass value --which is estimated considering the effects of optical depth and self-scattering, as well as a varying density profile, dust size distribution, and temperature profile-- is between a factor of 4.5 and 5.9 higher than the mass one would estimate from the 0.87 mm or 1.3 mm fluxes of TW Hya following the standard approximations used in disk demographic studies. The main factor causing this difference is likely the high optical depth at 0.87-1.3 mm, given that our modeling indicates that the disk is optically thick up to $r\sim55$ au at this wavelength. This difference in mass measurements might partially explain the discrepancy between the mass of exoplanetary systems and the disk masses predicted by ALMA surveys, but the early locking of solid mass in planetesimals is still a likely possibility.

\begin{acknowledgements}
We thank Anibal Sierra for helpful discussions. This work was supported by UNAM DGAPA-PAPIIT grants IN10818 and IG101321 and CONACyT Ciencia de Frontera grant number 86372. Support for this work was provided by NASA through the NASA Hubble Fellowship grant \#HST-HF2-51460.001-A awarded by the Space Telescope Science Institute, which is operated by the Association of Universities for Research in Astronomy, Inc., for NASA, under contract NAS5-26555.
This paper makes use of the following ALMA data: ADS/JAO.ALMA\#2015.1.00686.S, JAO.ALMA\#2015.1.00845.S, JAO.ALMA\#2015.A.00005.S, JAO.ALMA\#2016.1.00173.S, JAO.ALMA\#2016.1.00229.S,  JAO.ALMA\#2016.1.00629.S, JAO.ALMA\#2016.1.00842.S, JAO. ALMA\#2016.1.01158.S, JAO.ALMA\#2017.1.00520.S, and JAO.ALMA\#2018.1.01218.S. ALMA is a partnership of ESO (representing its member states), NSF (USA), and NINS (Japan), together with NRC (Canada), MOST and ASIAA (Taiwan), and KASI (Republic of Korea), in cooperation with the Republic of Chile. The Joint ALMA Observatory is operated by ESO, AUI/NRAO and NAOJ. This work has made use of data from the European Space Agency (ESA) mission {\it Gaia} (\url{https://www.cosmos.esa.int/gaia}), processed by the {\it Gaia} Data Processing and Analysis Consortium (DPAC, \url{https://www.cosmos.esa.int/web/gaia/dpac/} consortium). Funding for the DPAC has been provided by national institutions, in particular the institutions participating in the {\it Gaia} Multilateral Agreement.
\end{acknowledgements}

\begin{appendix}

\section{Modeling results for $a_{max}<200~\mu$m}

\begin{figure*}
\centering
\includegraphics[width=\textwidth]{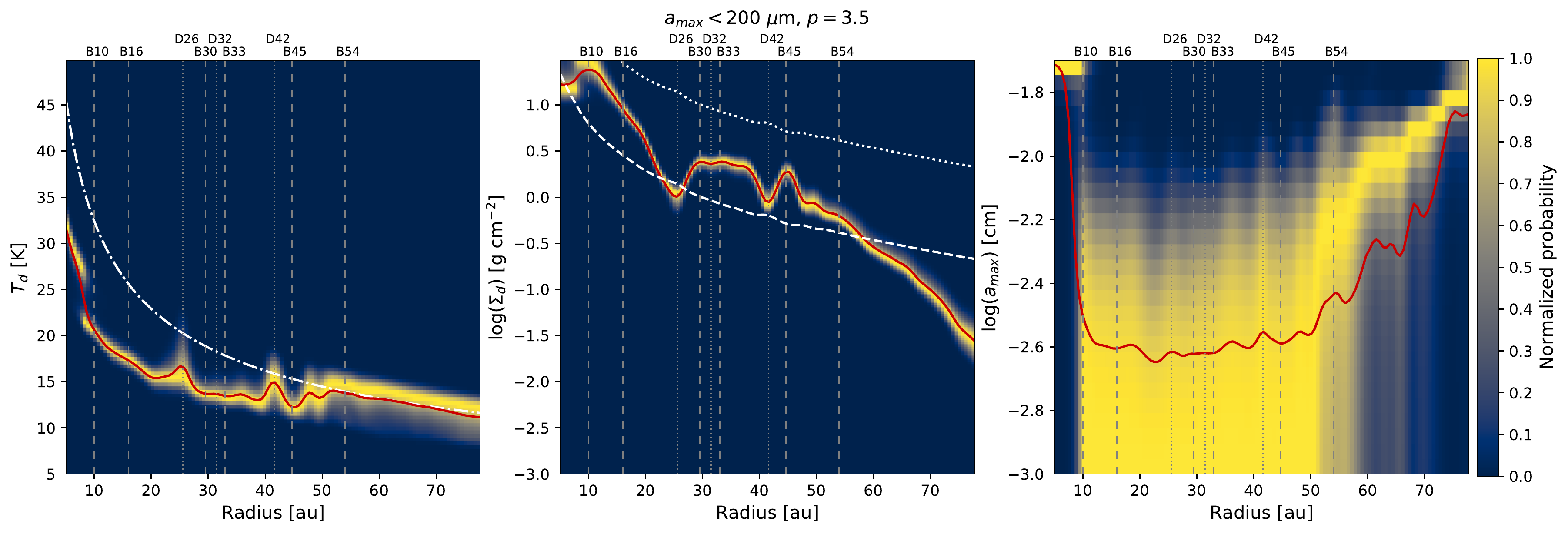}
\caption{Posterior probability distributions of the model parameters fitted to the multiwavelength observations, focusing on the family of solutions with smaller maximum grain sizes. The figure shows the same as Fig. \ref{fig:modlargedust} but for $a_{max}<200~\mu$m. }
\label{fig:modsmalldust_p3p5}%
\end{figure*}

\begin{figure*}
\centering
\includegraphics[width=0.8\textwidth]{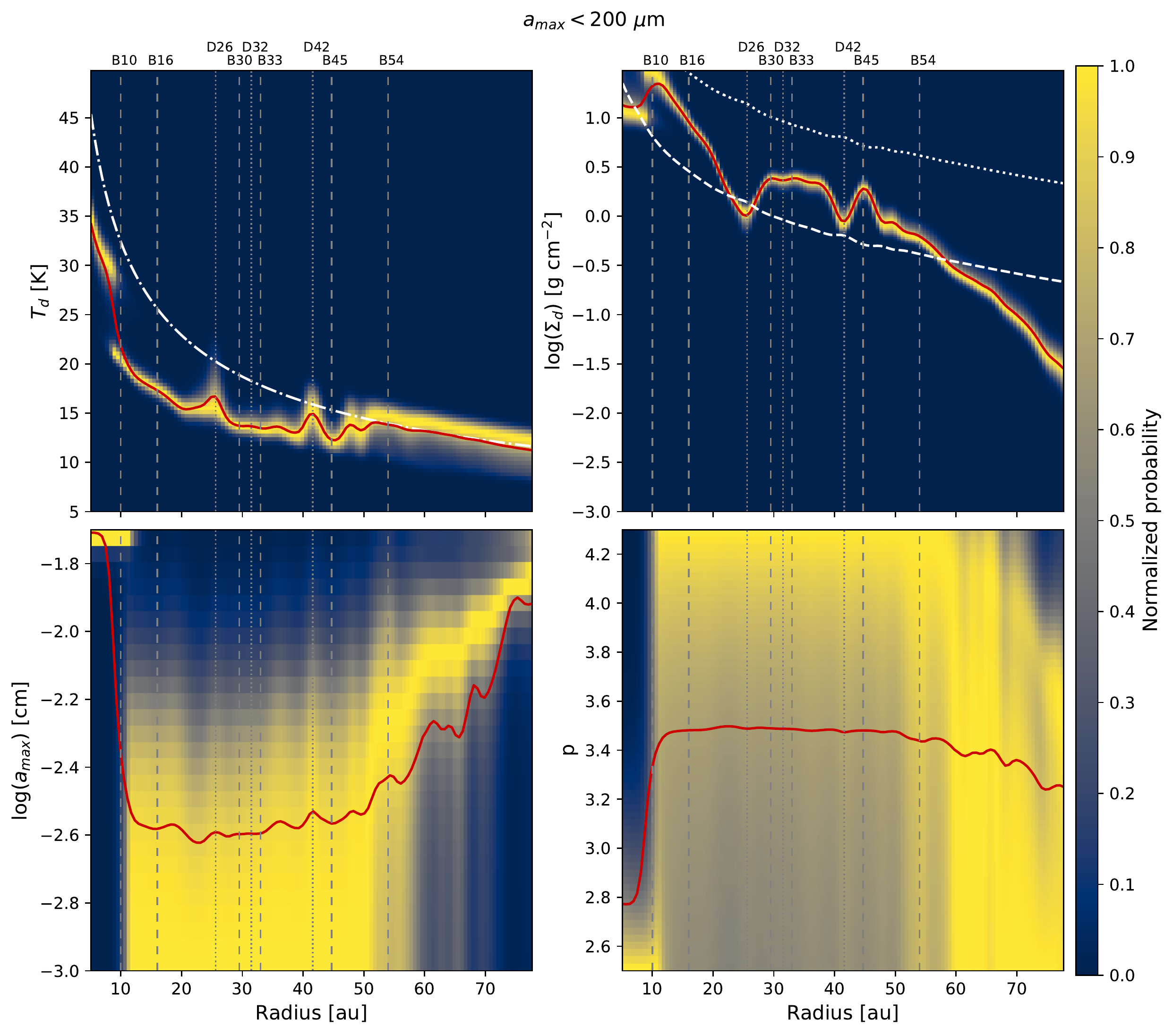}
\caption{Same as Figure \ref{fig:modsmalldust_p3p5}, but for the model with $p$ as a free parameter.}
\label{fig:modsmalldust}%
\end{figure*}

\begin{figure*}
\centering
\includegraphics[width=0.9\textwidth]{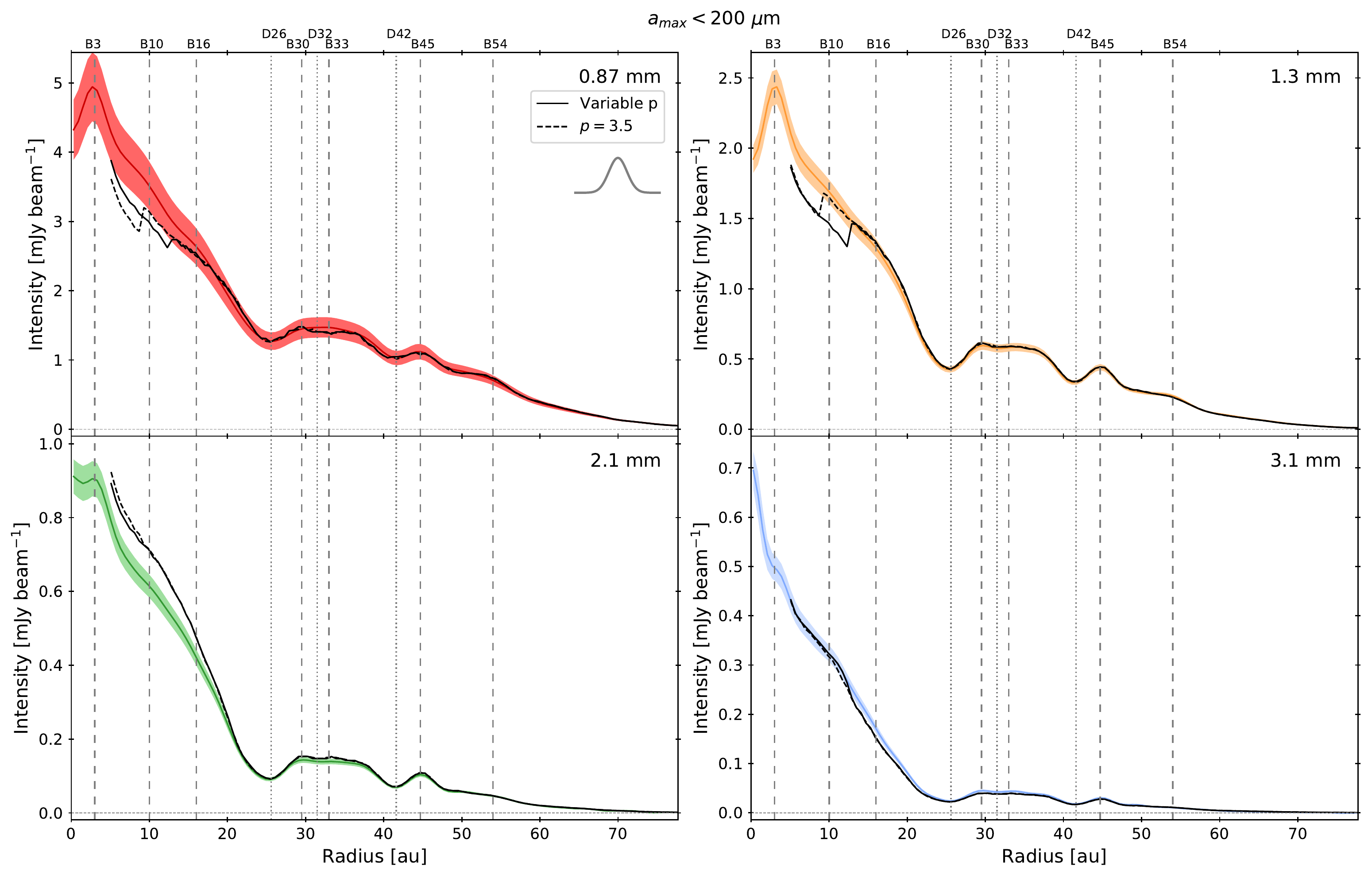}
\caption{Observed intensity profiles and model profiles with maximum posterior probability, focusing on the family of solutions with smaller maximum grain sizes. The lines and colors are the same as Fig. \ref{fig:maxlikeprofileb} but for $a_{max}<200~\mu$m.}
\label{fig:maxlikeprofiles}%
\end{figure*}

Figures \ref{fig:modsmalldust_p3p5} and \ref{fig:modsmalldust} show the radial marginalized posteriors of our dust characterization modeling for $a_{max}<200~\mu$m, fixing $p=3.5$ or including $p$ as a free parameter in the model, respectively. Additionally, Figure \ref{fig:maxlikeprofiles} shows the observed intensity profiles and model profile with maximum posterior probability for this same range of $a_{max}$. As mentioned above, we interpret these solutions as physically unrealistic, but show them here for completeness.

\section{Corner plots}

\begin{figure}
\centering
\includegraphics[width=0.5\textwidth]{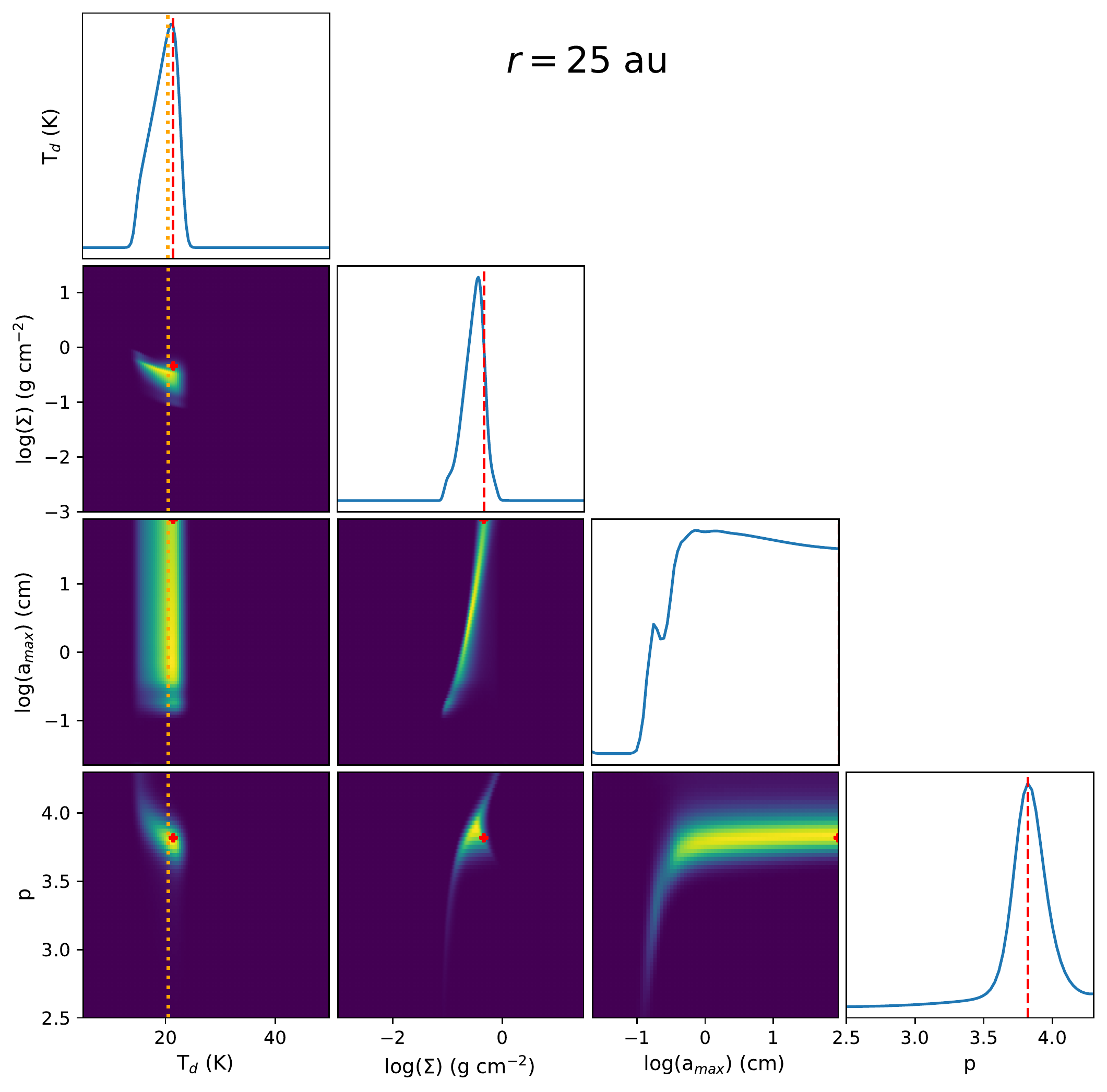}
\caption{Marginalized 2D posterior probability distribution between the different pairs of parameters in our $a_{max}>200~\mu$m model at the position of the D25 gap ($r=25$ au). The red dots and vertical lines indicate the values of the parameters at the maximum of the posterior probability distribution (not marginalized). The vertical dotted orange line indicates the central value of the temperature prior, obtained from equation \ref{eq:temp_prof}.}
\label{fig:corner1}%
\end{figure}

\begin{figure}
\centering
\includegraphics[width=0.5\textwidth]{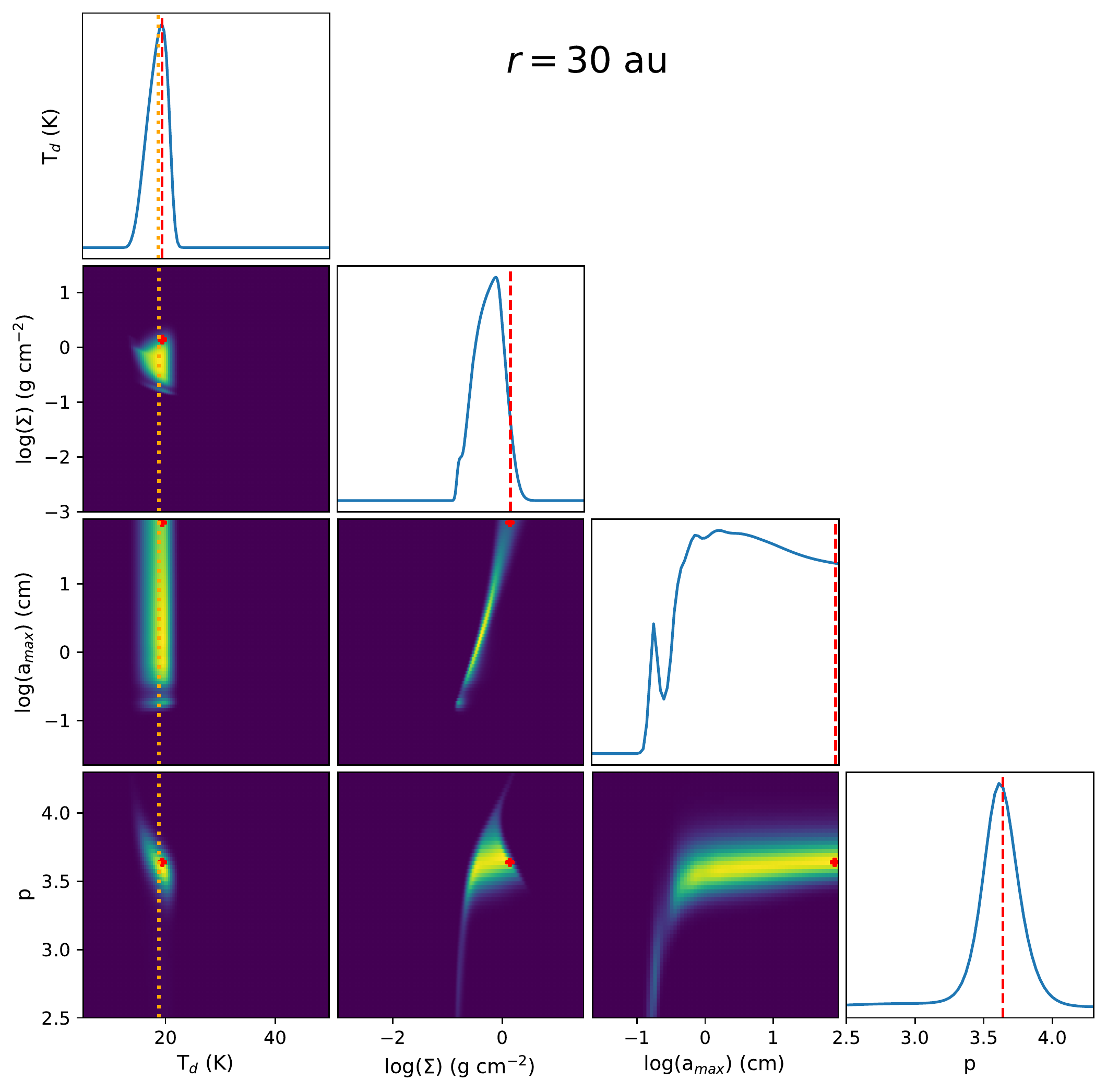}
\caption{Same as Figure \ref{fig:corner1}, but at the position of B30 ring ($r=30$ au).}
\label{fig:corner2}%
\end{figure}

\begin{figure}
\centering
\includegraphics[width=0.5\textwidth]{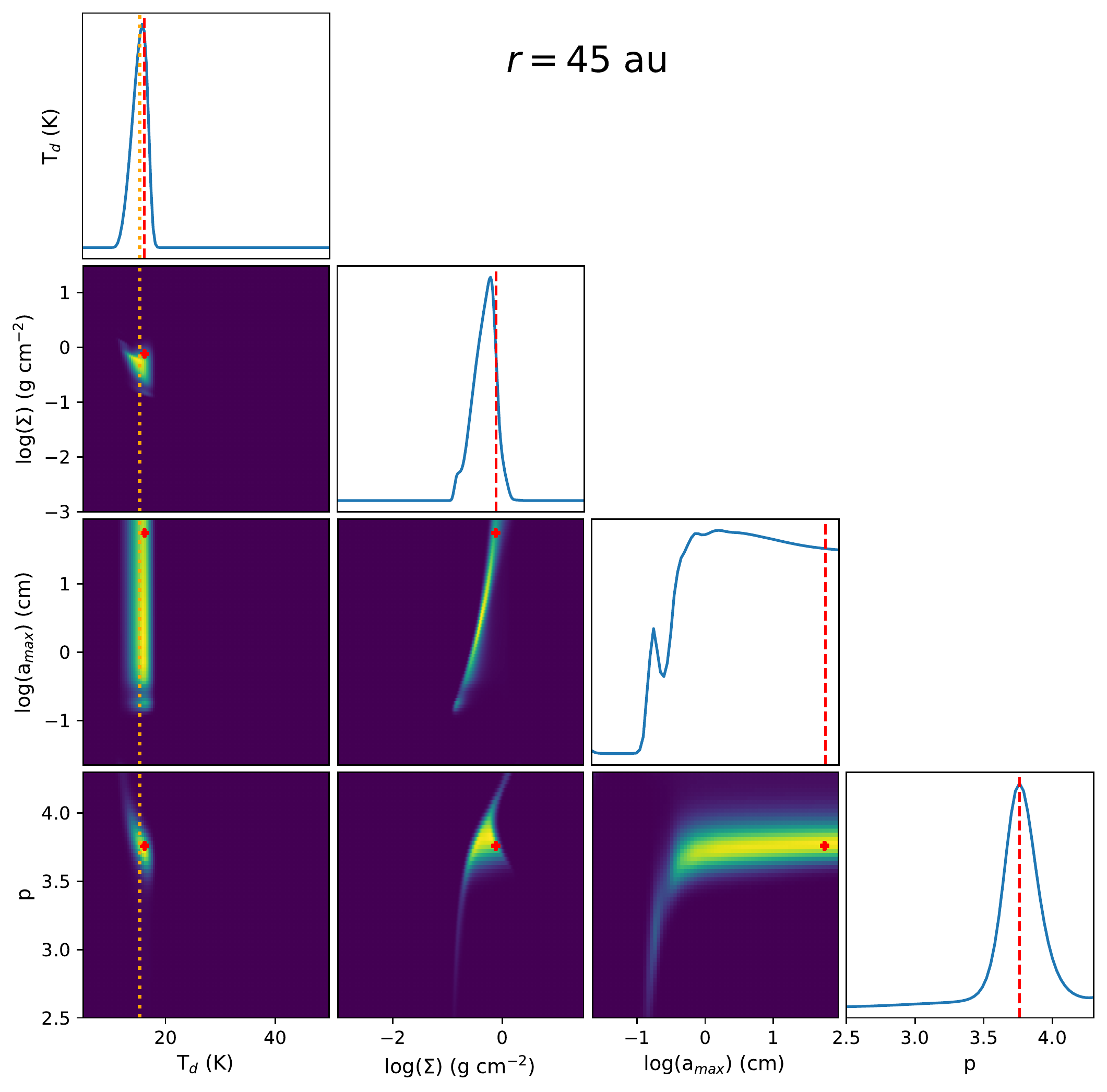}
\caption{Same as Figure \ref{fig:corner1}, but at the position of B45 ring ($r=45$ au).}
\label{fig:corner3}%
\end{figure}

We show the marginalized 2D posterior probability distribution between the different pairs of parameters in our model at some specific radii in the cornerplots in Figures \ref{fig:corner1}, \ref{fig:corner2}, and \ref{fig:corner3}. These plots are obtained from the results with $a_{max}>200~\mu$m of our model with $p$ as a free parameter, and at the radii of 25 au (D25 gap), 30 au (B30 ring), and 45 au (B45 ring).

\end{appendix}

\end{document}